\documentclass[usenatbib]{mn2e}
\def\kms{\,{\rm km}\,{\rm s}^{-1}}
\def\hmpc{\,{h^{-1} {\rm Mpc}}}
\def\mpch{\,{h {\rm Mpc}^{-1}}}

\def\ph#1{\phantom{#1}}

\usepackage{epsfig}

\usepackage{amssymb}

\usepackage{amsmath}

\begin{document}




\title[Cosmic Bulk Flow]{Cosmic bulk flows on 50\,\textit{h}$^{-1}$Mpc scales: A
Bayesian hyper-parameter method and multishell likelihood
analysis}

\author[Ma \& Scott]{Yin-Zhe Ma$^{1,2,\dagger}$ \& Douglas Scott$^{1,\star}$\\
$^1$Department of Physics and Astronomy, University of British Columbia, Vancouver, V6T 1Z1, BC Canada.\\
$^2$Canadian Institute for Theoretical Astrophysics, Toronto, M5S 3H8, Ontario, Canada.\\
emails: $^{\dagger}$mayinzhe@phas.ubc.ca;\,
$^{\star}$dscott@phas.ubc.ca}

\maketitle

\begin{abstract}
It has been argued recently that the galaxy peculiar velocity
field provides evidence of excessive power on scales of $50\hmpc$,
which seems to be inconsistent with the standard $\Lambda$CDM
cosmological model. We discuss several assumptions and conventions
used in studies of the large-scale bulk flow to check whether this
claim is robust under a variety of conditions. Rather than using a
composite catalogue we select samples from the SN, ENEAR, SFI++
and A1SN catalogues, and correct for Malmquist bias in each
according to the \textit{IRAS\/} PSCz density field. We also use
slightly different assumptions about the small-scale velocity
dispersion and the parameterisation of the matter power spectrum
when calculating the variance of the bulk flow. By combining the
likelihood of individual catalogues using a Bayesian
hyper-parameter method, we find that the joint likelihood of the
amplitude parameter gives $\sigma_8=0.65^{+0.47}_{-0.35}$ (68 per
cent confidence region), which is entirely consistent with the
$\Lambda$CDM model. In addition, the bulk flow magnitude, $v \sim
310 \kms$, and direction, $(l,b)\sim (280^{\circ} \pm 8^{\circ},
5.1^{\circ} \pm 6^{\circ})$, found by each of the catalogues are
all consistent with each other, and with the bulk flow results
from most previous studies. Furthermore, the bulk flow velocities
in different shells of the surveys constrain ($\sigma_{8}$,
$\Omega_{\rm{m}}$) to be ($1.01^{+0.26}_{-0.20},
0.31^{+0.28}_{-0.14}$), for SFI++ and ($1.04^{+0.32}_{-0.24},
0.28^{+0.30}_{-0.14}$) for ENEAR, which are consistent with {\it
WMAP\/} 7-year best-fit values. We finally discuss the differences
between our conclusions and those of the studies claiming the
largest bulk flows.
\end{abstract}

\begin{keywords}
methods: statistical--galaxies: kinematics and dynamics --distance
scale--large-scale structure of Universe

\end{keywords}

\section{Introduction}
\label{vel_intro} The cosmic bulk flow is the streaming motion of
the galaxies surrounding our Milky Way system, due to the
gravitational pull of cosmic structure on large scales. In the
gravitational instability paradigm, for a galaxy at position
$\mathbf{r}$, the peculiar velocity of an individual galaxy at
time $t$ is given by \citep{Peebles93}
\begin{equation}
\mathbf{v}(\mathbf{r,}t)=\frac{\Omega_{\rm{m}}^{0.55}H_{0}}{4\pi }
 \int d^{3} \mathbf{r^{\prime }}\delta_{m}(\mathbf{r^{\prime },}t)
 \frac{\mathbf{r}- \mathbf{r^{\prime }}}
 {|\mathbf{r}-\mathbf{r^{\prime }}|^{3}},
\label{pecu_def}
\end{equation}%
where $\delta_{\rm m}(\mathbf{r})=(\rho (\mathbf{r})-\overline{\rho })/%
\overline{\rho }$ is the density contrast at position
$\mathbf{r}$, $\Omega_{\rm{m}}$ is the fractional matter density,
and $H_{0}$ is the Hubble constant.  The bulk flow is normally
considered as an average over a sufficiently large volume, with
some window function $w(r,R)$, so that the above linear
perturbation theory is applicable.  This average is defined as
\citep{Juszkiewicz90,Nusser11a}
\begin{equation}
\mathbf{V}_{\rm bulk}(\mathbf{r,}t)
 =\frac{\int d^{3}\mathbf{r}^{\prime }\mathbf{v}(\mathbf{r}^{\prime}
 \mathbf{,}t)w(\left\vert \mathbf{r}^{\prime }-\mathbf{r}\right\vert ,R)}
 {\int d^{3}\mathbf{r}^{\prime }w(\left\vert \mathbf{r}^{\prime }
 -\mathbf{r}\right\vert ,R)},  \label{bulk_def}
\end{equation}
where $\mathbf{v}(\mathbf{r,}t)$ is the 3-D peculiar velocity
field at time $t$, defined in Eq.~(\ref{pecu_def}). A complete
investigation of bulk flows of nearby galaxies should measure the
individual velocities of galaxies all over the observed volume.
However, realistic observational techniques, such as the
Tully-Fisher relation, only allow us to probe the radial component
of the peculiar velocities of galaxies. In addition, most of the
current observations can only cover a patch of sky with limited
depth, leading to large uncertainties when interpreting the
results.

Of course, none of these considerations are new.  There is already
a large literature on the study of the peculiar velocity field,
with particularly intense activity in the early 1990s \citep[see
overviews
in][]{Burstein,Courteau93,Latham91,CosmicVF,Strauss95,CosmicFlows}.
Investigating the relationship between velocities and densities
has great potential for constraining cosmological parameters, and
testing theories of gravity on large scales.  However, it has long
been realised that the construction of appropriate catalogues is
difficult, and that systematic effects can easily overwhelm
statistical noise.

In attempting to overcome these observational limitations, there
have been significant recent efforts in the community to
reconstruct bulk flow moments from the limited data available, and
to test their consistency with the $\Lambda$CDM cosmology. One of
the important issues lies in determining the proper weighting for
individual galaxy velocities in a catalogue in order to obtain
streaming motions.  Some of the published studies, such as
\cite{Sarkar07} and \cite{Abate09}, focus on a weighting scheme
that produces the maximum likelihood estimate of the bulk flow
\citep[see also][]{Watkins09}, which can minimise the measurement
noise. However, this weighting depends on the particular survey
geometry and statistical properties, which leads to a large
uncertainty when interpreting the constraints from combined data
sets.

\cite{Watkins09} proposed another method of estimating the bulk
flow of galaxy peculiar velocities. They focused on the problem of
how realistic surveys can be used to reconstruct the bulk flow at
a given depth. They developed a minimum variance weighting method
\citep{Watkins09,Feldman10}, which minimises the variance between
the real data catalogue and the ideal survey, and they applied it
to combined catalogues of peculiar velocity surveys. Surprisingly,
they found a very large bulk flow on $50\hmpc$ scales ($v=407\pm
81 \kms$) towards $l=287^{\circ}\pm 9^{\circ}$, $b=8^{\circ}\pm
6^{\circ}$, which prefers a large amplitude of fluctuations
($\sigma_{8}$), inconsistent with the {\it WMAP\/} 5-year results
\citep{Komatsu09}. Subsequent work has discussed a possible
explanation for this large bulk flow related to pre-inflationary
isocurvature perturbations \citep{Ma11}.

Contradicting the claim in \cite{Watkins09}, \cite{Nusser11a}
developed a method termed the ASCE (All Space Constrained
Estimate) which reconstructs the bulk flow from an all-space 3-D
velocity field to match the inverse Tully-Fisher relation. By
applying this method, as well as the Maximum likelihood method
\citep{Abate09}, to the Spiral Field $I$-band Survey (SFI++
survey, \citealt{Springob07}) catalogue, \cite{Nusser11a} found
the bulk flow on a sphere of $40\hmpc$ radius to be $v=333 \pm 38
\kms$, towards ($l,b$)=($276^{\circ}\pm 3^{\circ} ,14^{\circ}\pm
3^{\circ}$), which is close to the results from the maximum
likelihood method. The estimated cosmological parameters, i.e.\
$(\Omega_{\rm{m}},\sigma_8)=(0.236,0.88)$, are consistent with the
$\Lambda$CDM model. However, since \cite{Nusser11a} only used the
SFI++ data set, it is still not clear whether it is the other data
sets used in \cite{Watkins09} which led to the significantly
different results.

Any analysis which claims to strongly rule out the simple
inflationary $\Lambda$CDM model deserves careful scrutiny, since a
confirmed discordance would have profound consequences for our
understanding of the large-scale Universe.  We can identify four
potential problems in \cite{Watkins09} which may potentially skew
the likelihood and bias the results. Firstly, the inhomogeneous
Malmquist bias is not corrected for in most catalogues, for
example: ENEAR \citep{Costa00,Bernardi02,Wenger03}; SN
\citep{Tonry03}; SC \citep{Giovanelli98}; EFAR \citep{Colless01};
and Willick \citep{Willick99}. This deficiency can significantly
bias the distance estimates.  Secondly, the distance errors from
the Tully-Fisher and Fundamental Plane methods can be comparable
to the measured velocities as the surveys go deeper, and moreover
a simple model of Gaussian errors is almost certainly
inappropriate as systematics come to dominate the distance
estimation. Therefore the velocity data beyond $100\hmpc$ become
both very noisy and unreliable in assessing the bulk flow.
Thirdly, directly combining various catalogues with different
calibration methods can also induce systematic errors and a
spurious flow. Finally, the assumption of a unique small scale
velocity dispersion $\sigma_{\ast}=150 \kms$ may be too small for
some of the surveys (e.g. SFI++ prefers $400 \kms$,
\citealt{Ma11}), perhaps skewing the constraints on the
cosmological parameters $\sigma_{8}$ and $\Omega_{\rm{m}}$. The
purpose of this paper is to investigate carefully the analysis
presented in \cite{Watkins09}, and to combine each catalogue with
a Bayesian hyper-parameter method to test for consistency with the
usual $\Lambda$CDM perturbation theory. As we have seen from
\cite{Nusser11a}, different statistical methods should not
dramatically alter the results, so we will focus on the `minimal
variance' scheme \citep{Watkins09,Feldman10}.

A further motivation for this paper is as an extension to the
velocity-gravity comparison work we have already carried out in
\cite{Ma12}. In that paper we compare the observational peculiar
velocity data with the reconstructed velocity field from the {\it
IRAS\/} PSCz catalogue, and fit the linear growth rate parameter,
$\beta$. In this new paper, we do not discuss the small-scale
modes, but will reconstruct the bulk motion of galaxies on
distances ${\sim}\,50\,h^{-1}$Mpc. We will perform a direct
comparison between observational data and $\Lambda$CDM model
predictions for the bulk flow velocity, and constrain the
cosmological parameters $\sigma_8$ and $\Omega_{\rm m}$. In
addition, we will extend the minimal variance scheme suggested in
\cite{Watkins09} and \cite{Feldman10} to a multishell likelihood
method. Furthermore, we will directly investigate the reason for
the apparently large flows found in \cite{Watkins09} and
\cite{Feldman10}.

This paper is organised as follows. We first list the data sets
used in Section~\ref{vel_data}, and then discuss the data
selection criterion in Section~\ref{dataselect}. For the selected
data, we correct the inhomogeneous Malmquist bias for the distance
estimate (Section~\ref{mbcorrect}). In Section~\ref{mv_scheme}, we
first illustrate how to quantify the variance of the bulk flow at
any particular depth (Section~\ref{mean_s_v}), then we review the
minimum variance weighting scheme proposed in \cite{Watkins09} to
measure the bulk flow at a given depth (Section~\ref{vel_mv}), and
furthermore we present the likelihood function for each individual
catalogue (Section~\ref{individual_like}) and the hyper-parameter
approach used to combine different data sets
(Section~\ref{vel_bayes_combine}). Then in
Section~\ref{result_discuss} we compare our findings with those in
\cite{Watkins09}.  We first confirm that we can accurately
reproduce the results in \cite{Watkins09} by adopting the same
conventions; then in Section~\ref{bfmoment} we show our
constraints on bulk flow moments by performing the full likelihood
analysis for each individual catalogue rather than the combined
catalogue. In Section~\ref{cosmologypara}, we apply the Bayesian
hyper-parameter method to combine the likelihoods of different
catalogues, in order to avoid the systematics that may affect the
constraints. This allows us to assess the consistency of each
individual catalogue, and to work out the cosmological parameters
in the combined likelihood. In Section~\ref{correlation-depth}, we
extend our likelihood analysis to consider bulk flows in multiple
shells in a survey, and their covariance matrix, and we compare
our findings with {\it WMAP\/} 7-year best-fit values and the
results from \cite{Nusser11a}. Our discussion and conclusion are
summarised in Section~\ref{vel_conclude}.

Note that although $H_0$ ($=100h\,{\rm km}\,{\rm s}^{-1}{\rm
Mpc}^{-1}$) is now determined with reasonable accuracy, throughout
this paper we continue to adopt the convention of giving distances
in units of $\hmpc$ for ease of comparison with previous results.

\section{Data}
\subsection{Catalogues}
\label{vel_data} We will use four different samples coming from
recent peculiar velocity surveys to reconstruct the bulk flow.
These samples are listed from the nearest to the most distant (see
also \citealt{Watkins09,Feldman10,Turnbull12}). Our four samples
consist of the ENEAR catalogue
\citep{Costa00,Bernardi02,Wenger03,Hudson94}, the SN catalogue
\citep{Tonry03}, the SFI++ catalogue \citep{Springob07} and the
A1SN catalogue \citep{Turnbull12,Jha07,Hicken09,Folatelli10}. For
detailed discussion and analyses of these four samples, including
their characteristic depths, typical distance errors and data
compilation, we refer readers to Section 3 of \citealt{Ma12}.

We should mention that in \cite{Watkins09} and \cite{Feldman10}
five other catalogues, namely SBF \citep{Tonry01}, SC
\citep{Giovanelli98,Dale99}, SMAC \citep{Hudson99,Hudson04}, EFAR
\citep{Colless01} and Willick \citep{Willick99}, were used to
reconstruct the bulk flow of galaxies. In contrast to the
previously described four catalogues, these samples are either
very distant and therefore have large errors, or very sparse in
which case the survey geometry is complicated. \cite{Watkins09}
combined these five low-quality catalogues with the previous four
higher quality catalogues to form a larger `COMPOSITE' catalogue,
and found an excess power of flow on scales of $50\hmpc$. However,
there is potential danger in combining various catalogues with
different calibration schemes. One concern is that the very
distant samples with large systematics may be inducing a spurious
large-scale flow.

In order to investigate this we tried to reproduce the `excess
flow' effect by using the suspicious COMPOSITE catalogue (see
Section~\ref{mv_scheme}). However, in the subsequent more careful
analysis, we will only use the ENEAR, SN, A1SN and SFI++
catalogues, with the following data selection criterion and
Malmquist bias correction.

\subsection{Data selection}
\label{dataselect} We listed four different peculiar velocity
catalogues in Section~\ref{vel_data}. In these catalogues, the
samples beyond the $80\hmpc$ scale are also quite sparse and
suffer from large errors due to uncertainties in the distance
indicators; therefore we trim the data sets at $80\hmpc$ in order
to reconstruct the bulk flow moments accurately on the $50\hmpc$
scale.

In addition, since some of the samples in the SFI++ catalogue with
$d \lesssim 30\hmpc$ are affected by localised non-linear
structures, giving very large velocities \citep{Ma12}, we excluded
these high velocity samples ($|v|>3000 \kms$) from the SFI++
catalogue. The classification of the data in each catalogue is
listed in Table~\ref{tab1}.

\begin{table}
\begin{centering}
\begin{tabular}{@{}lcc}
\hline
 & $d \leq 80$ & $80 \leq d \leq 200$\\
\hline \noalign{\vspace{-3pt}}
\ ENEAR & $\ph{0}669$ & $\ph{00}28$\\
\ SN & $\ph{00}78$ & $\ph{00}25$\\
\ SFI++ & $2404$ & $1052$\\
\ A1SN & $\ph{0}153$ & $\ph{00}92$\\
\noalign{\vspace{-3pt}} \hline
\end{tabular}%
\caption{Peculiar velocity samples. The two columns give the
number of galaxies within the range $d \leq 80\hmpc$ (used in this
paper) and $80 < d < 200\hmpc$ (considered as outliers).}
\label{tab1}
\end{centering}
\end{table}

\subsection{Malmquist bias correction}
\label{mbcorrect}

\begin{figure*}
\centerline{\includegraphics[bb=0 0 508
342,width=3.0in]{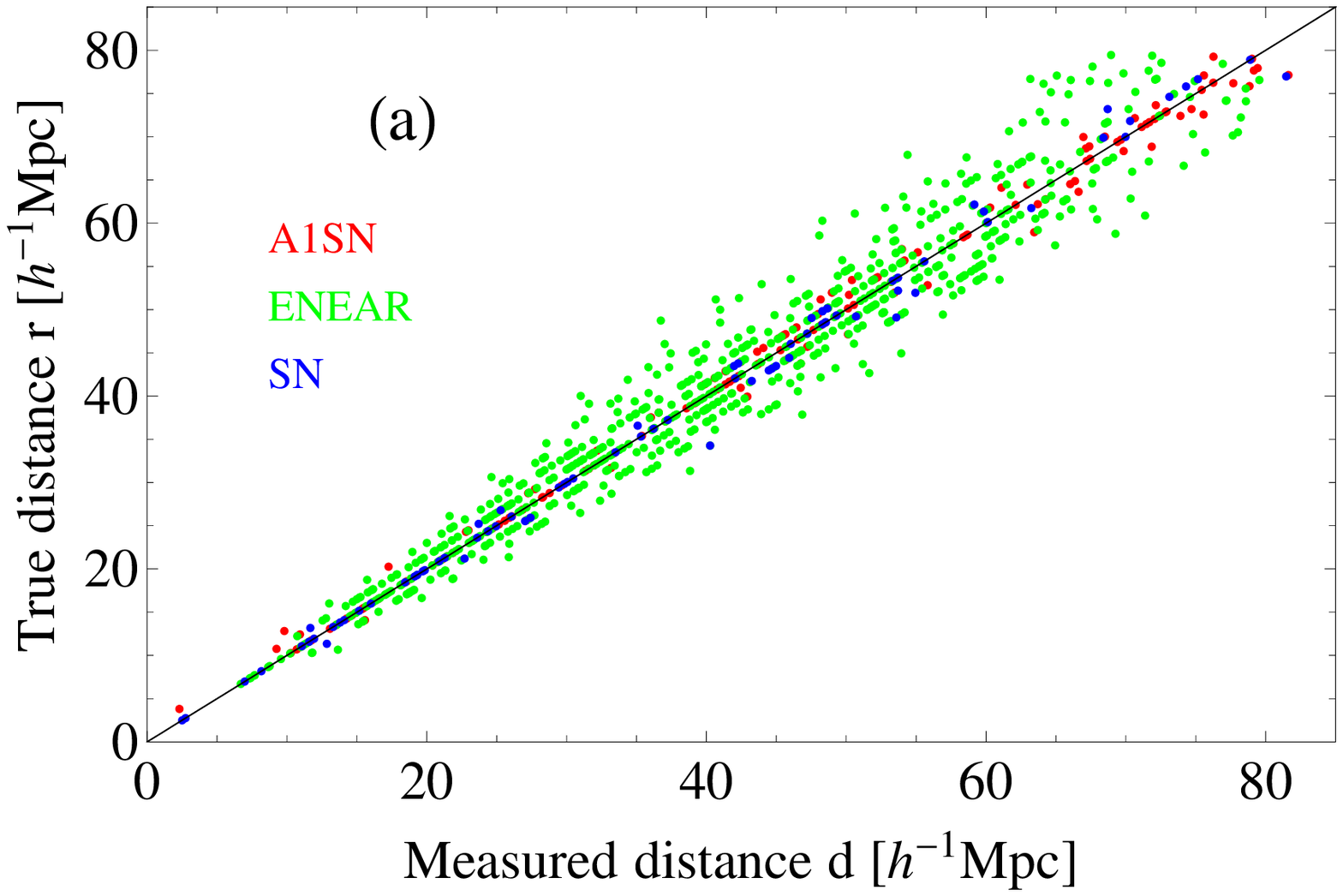}
\includegraphics[bb=0 0 517 327, width=3.2in]{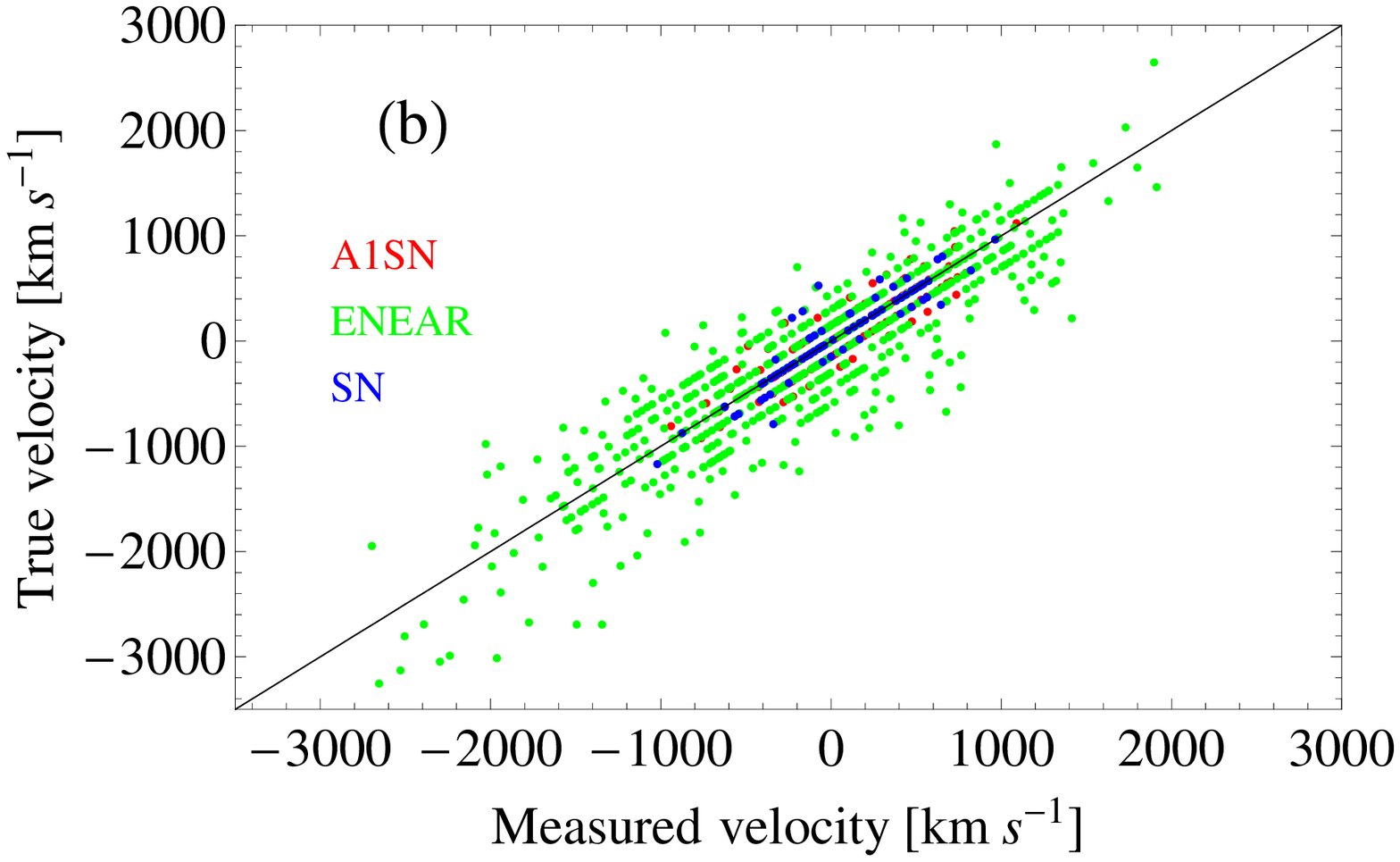}}
\caption{Inhomogeneous Malmquist bias correction. The $n(r)$
function of Eq.~(\ref{MBP1}) is interpolated by using
\textit{IRAS\/} PSCz density samples.} \label{MBcorrect}
\end{figure*}

In the above description of velocity catalogues, two different
distance indicators, the Tully-Fisher relation and the Fundamental
Plane method, have been used for determining the SFI++ and ENEAR
distances. In addition, supernova luminosities are used in
calibrating the distance of the SN and A1SN catalogues.

The large scatter of distance indicators suggests that objects
with inferred distance $d$, may come from a wide range of possible
true distances. The effect usually referred to as Malmquist bias
\citep{Malmquist20,Hendry93} is related to the probability
distribution of true distance $r$, given the measured distance $d$
with its measurement error. The desired function is
\citep{Lynden-Bell88a,Strauss95}
\begin{eqnarray}
P(r|d)=\frac{r^{2}n(r)\exp\left(- \frac{[\ln(r/d)]^2}{2
\Delta^{2}} \right)}
 {\int^{\infty}_{0} dr\, r^{2}n(r)
 \exp\left(- \frac{[\ln(r/d)]^2}{2
\Delta^{2}} \right)}, \label{MBP1}
\end{eqnarray}
where $n(r)$ is the radial density distribution, and
$\Delta=(\ln(10)/5)\sigma \simeq 0.46 \sigma$ is the fractional
distance uncertainty of distance indicators. Note that for the
Tully-Fisher and Fundamental Plane methods, the typical errors are
around $20$ per cent, and for Type Ia supernovae, the typical
error is around $6$--$8$ per cent.

The simplest case is \textit{homogeneous Malmquist bias}
\citep{Strauss95,Hudson94}, in which, the number density is
constant, so that Eq.~(\ref{MBP1}) becomes independent of density,
with
\begin{eqnarray}
P(r|d)=\frac{1}{\sqrt{2\pi}(d\Delta)}
\exp\left(-\frac{9}{2}\Delta^2 \right) \left[x^{2}
\exp{\left(-\frac{[\ln x]^2}{2 \Delta^2}\right)}\right],
\end{eqnarray}
where $x=r/d$ is the ratio between the true distance and the
measured distance. One can verify that the expectation of $r$,
$E(r|d)=\int r P(r|d)\rm{d}r =d \rm{e}^{7\Delta^2/2}$. This means
that even for a constant density distribution of galaxies, the
distance indicator is still generally biased. This is due to the
fact that, near the measured distance $d$, there are more galaxies
in shells of larger distance than smaller distance, so it is more
probable that the true distance is greater than the measured
distance i.e.\ $E(r|d)>d$.

However, in the more general case, the gradient of the number
density is not negligible, and this either reinforces or works
against the volume effect -- \textit{inhomogeneous Malmquist
bias}. If the gradient of the number density is positive, there
will be even more galaxies at the larger distances than in the
constant density case, i.e.\ the inhomogeneity reinforces the
homogeneous Malmquist bias; on the other hand, if the gradient is
negative, then the effective is opposite, and the inhomogeneous
Malmquist bias partially cancels the homogeneous Malmquist bias
effect.

To quantify the inhomogeneous Malmquist bias correctly, we use the
real-space reconstructed positions of the PSCz galaxies as mass
tracers to interpolate the mass density field on a cubic grid of
Length $192 \hmpc$ and mesh size $1.5 \hmpc$, smoothed with a
Gaussian filter of $5 \hmpc$. The field on the lattice is then
interpolated along the line of sight to each object in the
catalogue. The value of $n(r)$ along the line of sight is
specified at the position of $21$ equally-spaced points, with a
binning of $1.5 \hmpc$. Finally, Eq.~(\ref{MBP1}) is used to
predict $r$ from $d$ using a Monte Carlo rejection procedure.

We re-examine the catalogues described in Section~\ref{vel_data},
and correct for Malmquist bias according to Eq.~(\ref{MBP1}) for
the ENEAR, SN and A1SN samples\footnote{The SFI++ catalogue
\citep{Springob07} was already corrected for Malmquist bias.}. We
plot the measured distance (before Malmquist bias correction) and
corresponding true distance (after Malmquist bias correction) in
the left panel of Fig.~\ref{MBcorrect}. One can see that removing
the bias tends to place galaxies at larger distances, although the
shift is not very significant. The right panel of
Fig.~\ref{MBcorrect} shows the comparison between the uncorrected
and corrected line-of-sight peculiar velocities.

\section{Measuring the bulk flow}
\label{mv_scheme} For linear perturbation theory within the
$\Lambda$CDM paradigm, the velocity field at any spatial point
$\mathbf{v}(\mathbf{r},t)$ is directly related to the underlying
density field through Eq.~(\ref{pecu_def}). What we are interested
in is the bulk flow moment of the velocity field in a spherical
region. Therefore in this section, we first calculate the
mean-squared variance of the bulk flow (in Eq.~(\ref{bulk_def}))
which can be used to quantify the amplitude of the flow at
different depths. Then we will review the `minimum variance'
method for weighting the data sets on different scales.  Finally
we will present the likelihood function that can be used to
constrain cosmology with the measured bulk flow.

\subsection{Mean-squared velocity in a top-hat region}
\label{mean_s_v} Real surveys can only observe galaxies out to a
particular depth $R$, which means that the `window function' has a
sharp cut-off:
\begin{eqnarray}
 w(\left\vert \mathbf{r}^{\prime }-\mathbf{r}\right\vert ,R) &=&0,\text{ if }
 \left\vert \mathbf{r}^{\prime }-\mathbf{r}\right\vert >R;  \notag \\
 &=&1,\text{ if }\left\vert \mathbf{r}^{\prime }-\mathbf{r}\right\vert
 \leqslant R.
\end{eqnarray}
Therefore, by measuring only a galaxy sample within this sphere of
radius $R,$ one can calculate the `streaming motion' through
Eq.~(\ref{bulk_def}) by averaging the velocity within the sphere.
The mean-squared velocity of the spherical region within radius
$R$ is therefore (see also \citealt{Ma11b})
\begin{eqnarray} \left\langle \left\vert \mathbf{v}_{\rm
bulk}(t)\right\vert_{R}^{2}
  \right\rangle
 &=&\left\langle \mathbf{v}_{\rm bulk}(\vec{x},t)_{R}\cdot
  \mathbf{v}_{\rm bulk}(\vec{x},t)_{R}
  \right\rangle_{\vec{x} \text{ } \text{all space} }  \notag \\
&=&\left( \frac{3}{4\pi R^{3}}\right)^{2}\frac{(H
\Omega^{0.55}_{m} a(t))^{2}}{(2\pi )^{3}}
\int d^{3}\vec{k}\widetilde{w}^{2}(k,R)\frac{P(k)}{k^{2}}  \notag \\
&=&\frac{(3H \Omega_{\rm{m}}^{0.55} a(t))^{2}}{2\pi^{2}}\int P(k)
 \left( \frac{j_{1}(kR)}{kR}\right)^{2}dk.
\end{eqnarray}
At the present epoch $H=H_{0}$ and $a=1$, so today
\begin{equation}
\left\langle \left\vert \mathbf{v}_{\rm
bulk}\right\vert_{R}^{2}\right\rangle
 = \frac{(3H_{0}\Omega_{\rm{m}}^{0.55})^{2}}{2\pi^{2}}\int P(k)
 \left( \frac{j_{1}(kR)}{kR}\right)^{2}dk.  \label{vbulk1}
\end{equation}
We expect Eq.~(\ref{vbulk1}) to be useful for quantifying the
non-zero velocity fluctuations of our local surroundings. For the
{\it WMAP\/} 7-year cosmological parameters \citep{Komatsu11}, the
typical bulk flow magnitude on a scale of $50 \hmpc$ from
Eq.~(\ref{vbulk1}) is $310 \kms$. We will compare this theoretical
value with the measured velocity catalogues.

\subsection{Minimum variance scheme}
\label{vel_mv} Bulk flow estimates are essentially weighted
averages of the individual velocities in a galaxy survey
\citep{Watkins09}. Previous work, such as \cite{Abate09} and
\cite{Sarkar07}, focused on the estimate that minimises the
uncertainties due to measurement noise, i.e.\ the maximum
likelihood estimation scheme, but did not make any correction for
the survey geometry. Thus the maximum likelihood bulk flow is
obviously dependent on a given survey's particular geometry and
statistical properties. On the other hand, \cite{Watkins09} and
\cite{Feldman10} instead addressed the question of how peculiar
velocity data can be used to statistically estimate a more
specialised quantity, the bulk flow of an ideal, densely-sampled
survey with a given depth. They developed a `minimal variance'
weighting scheme which produces an estimate of the bulk flow at
any particular depth. They found an excess in the power of the
bulk flow on scales of $50\hmpc$, which seems to exceed the
$\Lambda$CDM predictions at the $3\sigma$ level. In the following,
we will first review the minimum variance weighting scheme
developed in \cite{Watkins09} and \cite{Feldman10} and then
present the likelihood function for cosmological parameters.

A realistic survey consists of $N$ objects on the sky having
position $\mathbf{r}_{i}$ and measured line-of-sight velocity
$S_{i}$, with measurement error $\sigma_{i}$. The measured
line-of-sight velocity is assumed to have the form
$S_{i}=v_{i}+\delta_{i}$, where $v_{i}$ is the galaxy
line-of-sight velocity in the matter rest frame, and $\delta_{i}$
is a superimposed Gaussian random motion with variance
$\sigma^2_{i}+\sigma^2_{\ast}$, where $\sigma_{\ast}$ accounts for
the 1-D small-scale velocity dispersion.

Given an idealised survey with bulk flow velocity $U_{p}$
($p=1,2,3$) at a particular depth $R$, we need to determine the
weight $w_{p,i}$ which makes the `linear compression'
\begin{equation}
u_{p}=\sum^{N}_{i=1}w_{p,i}S_{i},  \label{ucompress}
\end{equation}
give the closest approximation of $U_{p}$ \citep{Feldman10}.  At
the same time, the line-of-sight velocity at position
$\mathbf{r}_{i}$ should take the form $v_{i}=\sum_{p}
U_{p}(\mathbf{x}_{p}\cdot \mathbf{r}_{i})$. In order for the
estimator $u_{p}$ to give the correct amplitude of the velocity
$U_{p}$, i.e.\ $\langle u_{p} \rangle=U_{p}$, the weight function
$w_{p,i}$ has to satisfy the following constraint:

\begin{equation}
\sum_{i} w_{p,i}\left(\mathbf{x}_{q}\cdot \mathbf{r}_{i}
\right)=\delta_{pq}. \label{weightconstr}
\end{equation}
We can apply the Lagrange multiplier approach to minimise the
average variance $\langle (u_{p}-U_{p})^2 \rangle$, i.e.\ minimise
the following quantity \citep{Feldman10}
\begin{equation}
\langle \left(u_{p}-U_{p} \right)^2 \rangle + \sum_{q}
 \left( \sum_{i}w_{p,i}(\mathbf{x}_{q}\cdot \mathbf{r}_{i})-\delta_{p q} \right).
\end{equation}
By plugging in Eq.~(\ref{ucompress}), one can expand the first
term and obtain
\begin{eqnarray}
\langle U^{2}_{p} \rangle - 2 \sum_{i} w_{p,i} \langle S_{i}U_{p}
\rangle
 + \sum_{i,j}w_{p,i}w_{p,j}\langle S_{i} S_{j} \rangle  \notag \\
 + \sum_{q}\lambda_{p q}\left(\sum_{i}w_{p,i}
 (\mathbf{x}_{q}\cdot \mathbf{r}_{i})-\delta_{p q} \right).
\end{eqnarray}
In order to find the weight function $w_{p,i}$ that can minimise
the variance, we take the derivative of the above equation and
equate it to zero:
\begin{equation}
-2 \langle S_{i} U_{p} \rangle+2 \sum_{j}w_{p,j} \langle S_{i}
S_{j}
 \rangle + \sum_{q}\lambda_{pq}(\mathbf{x}_{q} \cdot \mathbf{r}_{i})=0.
\label{weight_func1}
\end{equation}
From Eq.~(\ref{weight_func1}), one can solve for the weight
function $w_{p,i}$ as
\begin{equation}
w_{p,i}=\sum_{j}(G^{-1})_{ij}\left[ \left\langle
S_{j}U_{p}\right\rangle
 - \frac{1}{2}\lambda_{pq} (\mathbf{x}_{q} \cdot \mathbf{r}_{j}) \right],
\label{w_func1}
\end{equation}
where $G_{ij}=\langle S_{i}S_{j} \rangle$ is the covariance matrix
for the measured velocity. Since $S_{i}=v_{i}+\delta_{i}$ as
described above, one can write the covariance matrix $G$ as
\begin{eqnarray}
G_{ij} &=& \langle v_{i}v_{j} \rangle +
\delta_{ij}(\sigma^{2}_{\ast}+\sigma^{2}_{i})  \notag \\
&=& \langle (\hat{\mathbf{r}}_{i} \cdot
\mathbf{v}(\mathbf{r}_{i}))
 (\hat{\mathbf{r}}_{j} \cdot \mathbf{v}(\mathbf{r}_{j})) \rangle
 +\delta_{ij}(\sigma^{2}_{\ast}+\sigma^{2}_{i}) ,  \label{vel_Gnm1}
\end{eqnarray}
since $v_{i}$ and $\delta_{i}$ are not correlated. The first term
is the real space velocity correlation function, which is related
to the matter power spectrum in Fourier space,
\begin{equation}
\bigl\langle \bigl(\hat{\mathbf{r}}_{i} \cdot
\mathbf{v}(\mathbf{r}_{i})\bigr)
 \bigl(\hat{\mathbf{r}}_{j} \cdot \mathbf{v}(\mathbf{r}_{j})\bigr) \bigr\rangle
 =\frac{\Omega^{1.1}_{\rm m}\mathrm{\ }H^{2}_{0}}{2 \pi^{2}}\int \mathrm{\ } dk
 \mathrm{\ } P(k) \mathrm{\ } F_{ij}(k),  \label{vel_rnrm1}
\end{equation}
where the window function,
\begin{equation}
 F_{ij}(k)=\int \mathrm{\ }\frac{d^{2} \hat{k}}{4 \pi}
  \left(\hat{\mathbf{r}}_{i}\cdot \hat{\mathbf{k}} \right)
  \left(\hat{\mathbf{r}}_{j}\cdot \hat{\mathbf{k}} \right)
  \times \exp(i k \hat{\mathbf{k}}\cdot (\mathbf{r}_{i}
  -\mathbf{r}_{j})),  \label{vel_fmn_win}
\end{equation}
can be calculated analytically \citep{Ma11}. 

The correlation term $\langle S_{j}U_{p} \rangle$ is the average
product of measured velocity $S_{j}$ with the ideal bulk flow
moment $U_{p}$. The ideal bulk flow moment $U_{p}$ is the average
of the random velocities in an isotropic survey region. We assume
that the survey is a spherical region with radius $R$, therefore
we generate $N'=10^4$ random velocities in the top-hat region $R$
and calculate $\langle S_{j}U_{p} \rangle $ as
\begin{equation}
 \langle S_{j}U_{p} \rangle= \frac{1}{N^{\prime }}
  \sum^{N^{\prime}}_{n^{\prime }=1}
  \left( \mathbf{x}_{p} \cdot \mathbf{r}_{n^{\prime }}\right)
  \langle v_{n^{\prime }} v_{j} \rangle,  \label{vel_smup1}
\end{equation}
where the line-of-sight velocity correlation $\langle v_{n^{\prime
}} v_{j} \rangle$ can be calculated in the same manner as
Eq.~(\ref{vel_rnrm1}).

Therefore, the only unknown in Eq.~(\ref{w_func1}) is the Lagrange
multiplier matrix $\lambda$. We can plug Eq.~(\ref{w_func1}) into
the constraint equation (\ref{weightconstr}) to solve for
$\lambda_{pq}$:
\begin{equation}
\lambda_{pq}=\sum_{l}\left[ \sum_{ij}\left\langle
S_{j}U_{p}\right\rangle
G_{ij}^{-1}g_{l}(\mathbf{\hat{r}}_{j})-\delta_{pl}\right]
M_{lq}^{-1}, \label{vel_lambda_mat}
\end{equation}%
where the matrix $M$ is given by
\begin{equation}
M_{pq}=\frac{1}{2} \sum_{i,j} G^{-1}_{ij}(\mathbf{x}_{p}\cdot
 \mathbf{r}_{i})(\mathbf{x}_{q}\cdot \mathbf{r}_{j}).  \label{vel_Mmatrix}
\end{equation}
To summarise, by simulating an ideal survey at depth $R$ and using
Eqs.~(\ref{vel_Gnm1}), (\ref{vel_smup1}), (\ref{vel_lambda_mat})
and (\ref{vel_Mmatrix}), one can calculate the weight function
$w_{p,i}$ in Eq.~(\ref{w_func1}) and hence obtain the bulk flow
Moment at depth $R$ according to Eq.~(\ref{ucompress}).

\begin{figure*}
\centerline{\includegraphics[bb=0 0 602
395,width=3.2in]{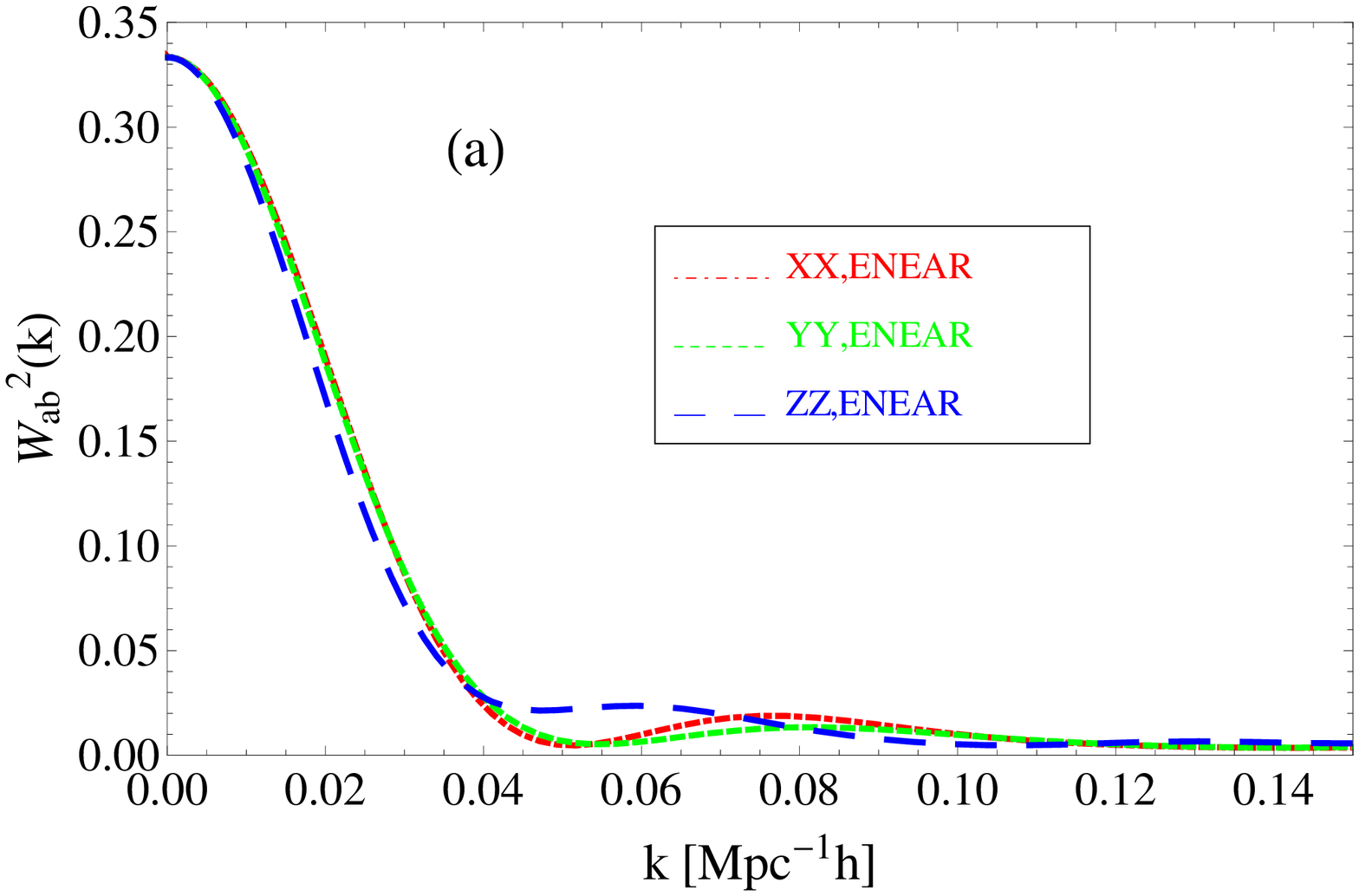}
\includegraphics[bb=0 0 638 422,width=3.2in]{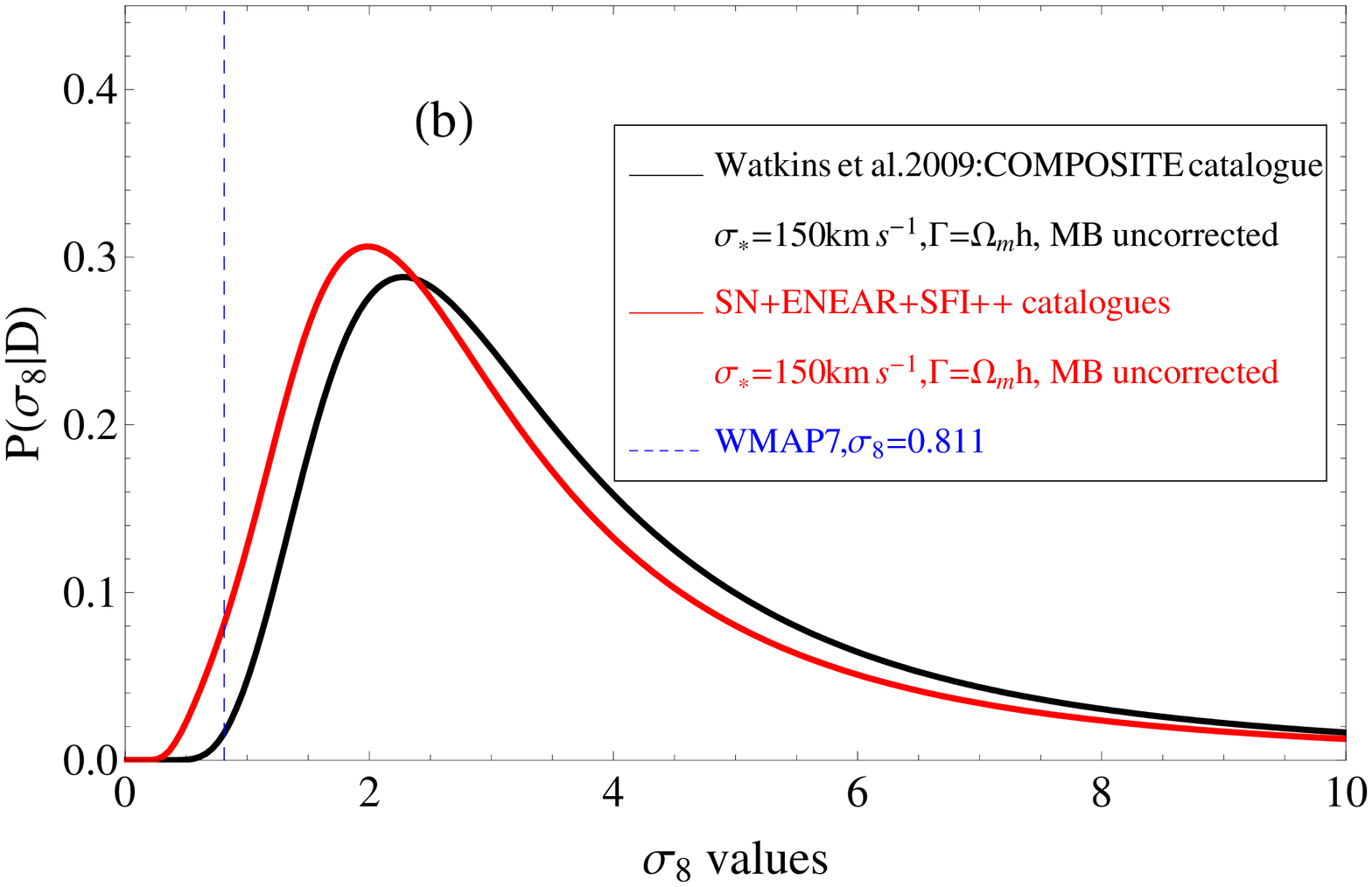}}
\caption{(a) Diagonal term of the window function $W^{2}_{ab}(k)$
(Eq.~(\ref{vel_w2pq})) for the ENEAR catalogue at depth $50 \hmpc$
(see Section~\ref{vel_data} for a description of the catalogues).
(b) Marginalised distribution of $\sigma_{8}$ for the COMPOSITE
catalogue. The black line is obtained by adopting the conventions
used in Watkins et al. (2009): no Malmquist bias correction;
COMPOSITE catalogue (direct combination of SBF, SN, ENEAR, SFI++,
EFAR, SC, SMAC and Willick); small scale velocity dispersion
$\sigma_{\ast}=150 \kms$; and matter power spectrum parameter
$\Gamma=\Omega_{\rm{m}}h$. Note that this black line is produced
by our own code and it matches figure 7 of \citealt{Watkins09}
very well. The red line corresponds to the case where we use all
these same conventions, but remove the SBF, EFAR, SC, SMAC and
Willick catalogues.} \label{reproduce1}
\end{figure*}

Once we obtain the bulk flow moment from the minimum variance
weighting scheme, we can calculate the covariance matrix and
therefore perform a full statistical analysis. The covariance
matrix of $u_{p}$ becomes
\begin{eqnarray}
C_{pq} &=& \langle u_{p}u_{q} \rangle  \notag \\
&=& \sum_{i,j}w_{p,i}w_{q,j}\langle S_{i} S_{j} \rangle  \notag \\
&=&\sum_{i,j}w_{p,i}w_{q,j} G_{ij}, \label{covar1}
\end{eqnarray}
which can be broken down into an instrumental noise term
\begin{equation}
C^{\rm
n}_{pq}=\sum_{i}w_{p,i}w_{q,i}\left(\sigma^{2}_{i}+\sigma^{2}_{\ast}
\right),  \label{noise_mat1}
\end{equation}
and a cosmic variance term
\begin{equation}
C^{\rm v}_{pq}=\frac{\Omega^{1.1}_{\rm m} H^{2}_{0}}{2 \pi^{2}}
\int dk \text{ }P(k) \text{ }W^{2}_{pq}(k),
\end{equation}
where the angle-averaged window function is
\begin{equation}
W^{2}_{pq}(k)=\sum_{i,j}w_{p,i}w_{q,j}F_{ij}(k).  \label{vel_w2pq}
\end{equation}
This window function describes the scale in $k$-space that the
catalogue actually probes. As an example, we plot this window
function for the ENEAR catalogue at depth $50 \hmpc$ in
Fig.~\ref{reproduce1}a. As expected, it matches the window
function of ENEAR as shown in figure 3 in \cite{Watkins09}
perfectly well. The shape of the curves reveal that the window
function decays rapidly for $k$ beyond $0.05 \mpch$, therefore the
non-linear regime of the matter power spectrum does not contribute
to the bulk flow moment -- bulk flow moments reflect perturbations
on large scales, $k \lesssim 0.05 \mpch$.

\subsection{Likelihood of Individual catalogues}
\label{individual_like} Given the reconstructed bulk flow moment
at some depth $R$ and its covariance matrix $C_{ab}$
(\ref{covar1}), the likelihood of cosmological parameters
$\mathbf{\theta}$ is
\begin{equation}
\mathcal{L}(\mathbf{\theta})= \frac{1}{(2 \pi)^{\frac{3}{2}}
 \det(C(\mathbf{\theta}))^{\frac{1}{2}}}\exp
   \left(-\frac{1}{2} \sum^{3}_{p,q=1}u_{p}C(\mathbf{\theta})^{-1}_{pq}u_{q}
   \right). \label{like1}
\end{equation}
Since the major effect of the constraints is on the amplitude and
shape of the matter power spectrum, in the following analysis we
will only vary $\sigma_{8}$ and $\Omega_{\rm{m}}$, while fixing
the other parameters at the {\it WMAP\/} values. We compute the
bi-variate likelihood and marginalise one parameter to obtain the
1-D posteriori distribution of the other parameter.

In order to demonstrate the accurate reproduction of the results
in \cite{Watkins09}, we use the same set of {\it WMAP\/} 5-year
best-fit parameters \citep{Komatsu09}: $\Omega_{\rm b}=0.0441$;
$h=0.719$; $n_{\rm s}=0.963$; $\tau=0.087$; $A_{\rm s}=2.41 \times
10^{-9}$; plus small-scale velocity dispersion
$\sigma_{\ast}=150\kms$. We also use the approximation
$\Gamma=\Omega_{\rm{m}}h$ to calculate the matter power spectrum
(see Appendix \ref{powerspectrum_form} for comparison with the
numerical result), as is used in \cite{Watkins09}. In
Fig.~\ref{reproduce1} we show that we can accurately reproduce the
window function $W^{2}_{ab}(k)$ (see Eq.~(\ref{vel_w2pq})) and
marginalised distribution of $\sigma_{8}$, as shown by the black
line in Fig.~\ref{reproduce1}b. These curves are very close to
those shown as figures~3 and 7 of \cite{Watkins09}.

In addition, since we will not use the SBF, SC, SMAC, EFAR and
Willick catalogues here, we need to test whether this `removal' of
data can substantially change the result. To do this, we use the
rest of the data in the COMPOSITE catalogue, i.e.\ the combination
of the SN, ENEAR and SFI++ catalogues (4256 samples in total), and
keep all other conventions the same as in \cite{Watkins09} to
calculate the likelihood of $\sigma_8$, and we obtain the red line
in Fig.~\ref{reproduce1}b. Comparing with the black line, one can
see that the removal of these sparse and distant samples can move
the peak of the likelihood towards lower values, but a very high
value of $\sigma_8$ is still preferred compared with the {\it
WMAP\/} constraint (dashed line). Therefore, the excessive power
in the bulk flow on $50\hmpc$ scales is not completely driven by
the inclusion of five sparse and fairly noisy samples -- we need
to investigate further to understand the reasons.

We make the following adjustments to the model parameters in order
to precisely compare the velocity field prediction with the
observational data:

\begin{itemize}
\item we use {\it WMAP\/} 7-year best-fit cosmological parameters
\citep{Komatsu11} to compute our prediction, i.e.\ $\Omega_{\rm
b}=0.0455$, $h=0.704$, $n_{\rm s}=0.967$, $\tau=0.088$ and $A_{\rm
s}=2.43 \times 10^{-9}$;

\item we use the value $\sigma_{\ast} = 400 \kms$ for the
intrinsic velocity dispersion,\footnote{\cite{Turnbull12} used a
thermal noise $250 \kms$.} in order to compute the covariance
matrix (Eq.~(\ref{noise_mat1})), since \cite{Ma11} showed that
this value is preferred for most catalogues;

\item in the formula for $P(k)$, we use the numerical result from
CAMB \citep{camb} instead of $\Gamma =\Omega_{\rm{m}}h$ to compute
the power spectrum, with the difference between the numerical
result and the parameterisation $\Gamma =\Omega_{\rm{m}}h$ being
shown in Appendix~\ref{powerspectrum_form}.
\end{itemize}

\subsection{Combining catalogues: Bayesian hyper-parameter method}
\label{vel_bayes_combine} In \cite{Watkins09}, the combined
catalogue, referred to as `COMPOSITE' is used to reconstruct the
bulk flow and constrain the cosmological parameters $\sigma_{8}$
and $\Omega_{\rm{m}}$. They found an excess power in the bulk flow
on a scale of $50\hmpc$, which suggests a high value of
$\sigma_{8}$ compared with the constraints from {\it WMAP\/}
5-year results \citep{Komatsu09} -- see Fig.~\ref{reproduce1}b.

Directly combining a variety of catalogues with different
calibration methods and systematics may not be a precise way of
exploring the combined constraints. Another way of carrying out
the combination is to first compute the likelihood of individual
data sets, and then directly combine them by multiplication, i.e.\
\begin{equation}
\mathcal{L}_{\rm joint}(\theta )=\prod
\limits_{k}^{N}\mathcal{L}_{k}(\theta ), \label{direct_comb}
\end{equation}
where $N$ is the number of data sets. Such a procedure assumes
that the quoted observational random errors can be trusted, and
that the two (or more) $\chi^{2}$ statistics have equal weights,
so that
\begin{equation}
\chi_{\rm joint}^{2}=\sum_{k}\chi_{k}^{2}.  \label{joint_chi2}
\end{equation}
However, when combing different data sets, one often wants to
assign different weights to them.  \cite{Lahav00} describe an
approach \citep[see also][for an earlier application of the same
idea in astrophysics]{Press} using
\begin{equation}
\chi_{\rm joint}^{2}=\sum_{k}\alpha_{k}\chi_{k}^{2},
\label{hyper_chi2}
\end{equation}%
where the $\alpha_{k}s$ are `hyper-parameters', which are to be
evaluated in a Bayesian way. Here $\chi^{2}$ for each data set is
\begin{equation}
\chi^{2}=\sum_{i}\frac{[x_{i}^{\rm obs}-x_{i}^{\rm the}(\theta
)]^{2}}{\sigma_{i}^{2}},
\end{equation}
where the summation is over $N$ measurements and $\sigma_{i}$ is
the error for each data point. By multiplying $\chi^{2}$ by
$\alpha$, each error $\sigma_{i}$ effectively becomes
$\alpha^{-\frac{1}{2}}\sigma_{i}$ and therefore if an experiment
underestimates (or overestimates) the systematic errors, the
hyper-parameter can scale the error by using relative weights.
Indeed, the hyper-parameters are useful in assessing the relative
weight for each different experiment. This procedure gives an
objective diagnostic for revealing experiments with problematic
error estimates and which therefore deserve further investigation
of their systematic or random errors.

It is worth clarifying that, in principle, the systematic effects
could have two different components, either an overall
multiplicative factor for the velocities, or else an extra
contribution to the measurement noise. Although the former kind of
systematic effect would be appropriate for some other kinds of
data (particularly where the dominant uncertainty is a linear
calibration factor), the effect on the peculiar velocity field is
more complicated than this.  We therefore focus our attention on
modelling the systematics as an additional source of noise,
effectively giving a different weighting of the signal-to-noise of
each data set.  For a discussion of related issues in other
branches of astrophysics see for example \cite{Gull89} and
\cite{Stompor09}.

In the same spirit of assigning weights to each data set,
\cite{Hobson02} calculated the joint distribution of cosmological
parameters for multiple data sets, in which the weight assigned to
each is determined directly by its own statistical properties. The
weights are considered in a Bayesian context as a set of
hyper-parameters, which are then marginalised over in order to
recover the posterior distribution as a function only of the
cosmological parameters of interest. In the case of a Gaussian
likelihood function, this marginalisation can be calculated
analytically, and it is shown that the joint probability
distribution, $P(D|\theta)$, when applying the hyper-parameter
approach is
\begin{equation}
\mathcal{L}_{\rm joint}(\theta )=\prod\limits_{k=1}^{N}
 \frac{2\Gamma \left( \frac{n_{k}}{2}+1\right) }
 {\pi^{n_{k}/2}\left\vert V_{k}\right\vert^{1/2}}
 \left( \chi_{k}^{2}+2\right)^{-\left(\frac{n_{k}}{2}+1\right) },
  \label{hyper_comb}
\end{equation}
where $n_{k},V_{k}$ and $\chi_{k}^{2}$ are the number of data,
covariance matrix and $\chi^{2}$ for the $k$th data set. In our
approach, flat priors on $\sigma_{8}$ and $\Omega_{\rm{m}}$ are
assumed. We will use Eq.~(\ref{hyper_comb}) to explore the use of
different catalogues to constrain cosmological parameters.

Once the distribution of Eq.~(\ref{hyper_comb}) is calculated, the
hyper-parameter is already marginalised over, which means that it
automatically incorporates the relative weights between each data
set and combines them in an objective way. Therefore, rather than
using the COMPOSITE catalogue to constrain cosmology, we will
first investigate the individual likelihoods for each data set,
and use the joint distribution in Eq.~(\ref{hyper_comb}) to
combine these data sets.

\section{Results}
\label{result_discuss} In this section, we will perform two
different analyses separately. The first one, in
Section~\ref{bfmoment} and \ref{cosmologypara}, will focus on
reconstructing $v_{\rm{bulk}}$ on $50 \hmpc$ scales, and use it to
constrain cosmological parameters. In the second analysis, we will
extend the `minimal variance' scheme to consider the cumulative
bulk flow at different radii, and to explore the joint constraints
on cosmology from all the bulk flows in different shells.

\subsection{Bulk flow moments}
\label{bfmoment}
\begin{figure*}
\centerline{\includegraphics[bb=0 0 521 352,width=3.0in]{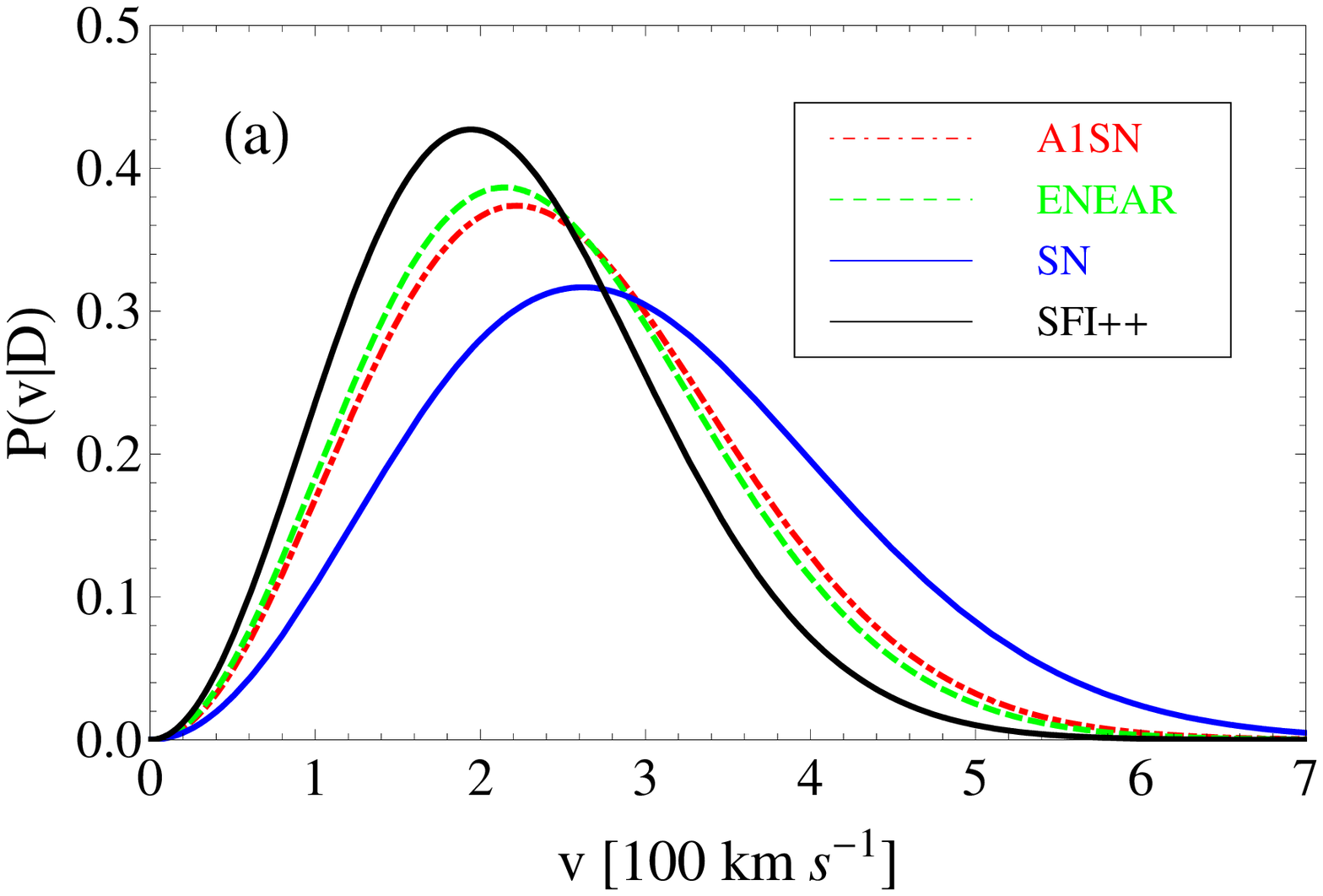}
\includegraphics[bb=0 0 464 472,width=2.8in]{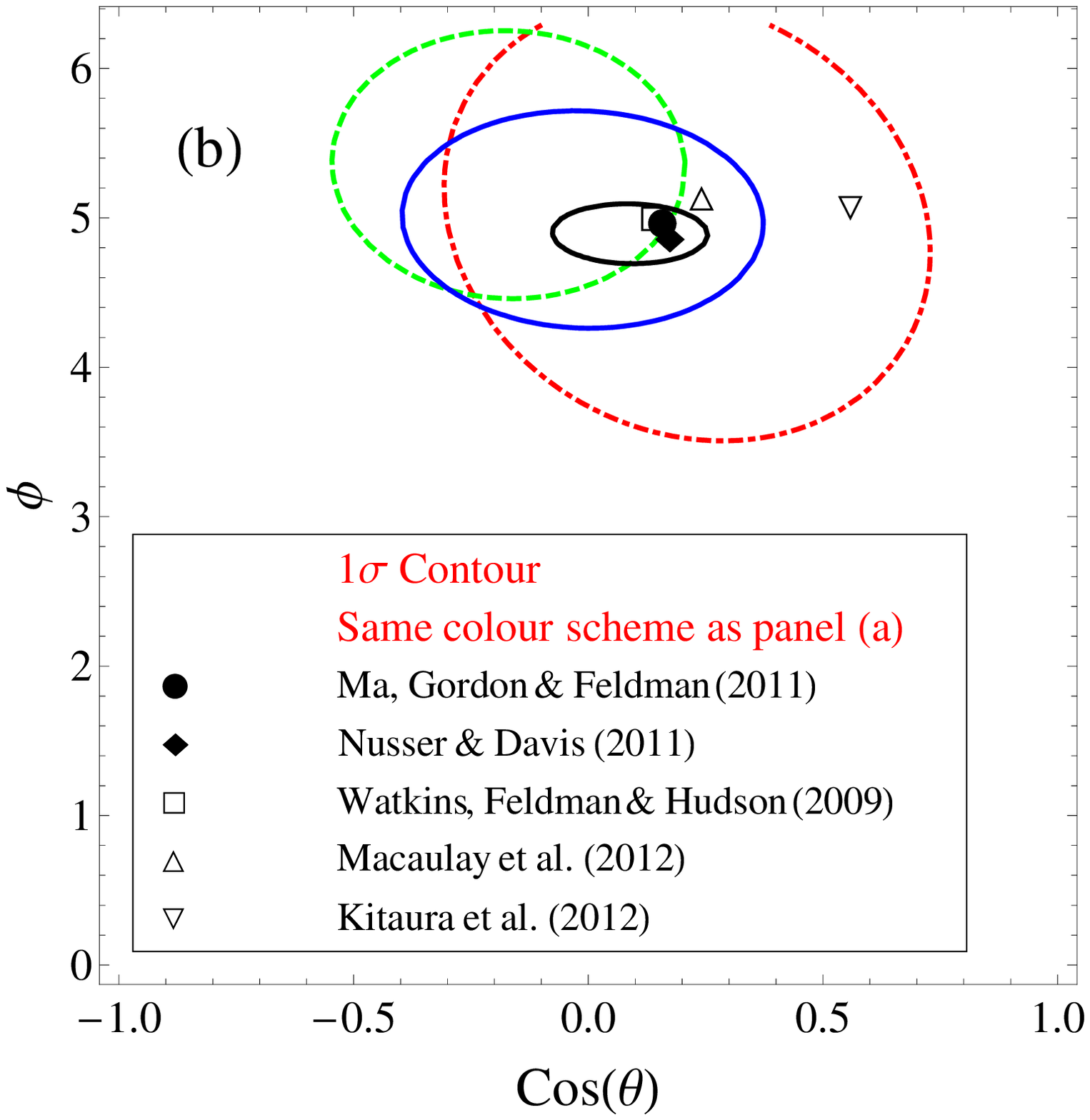}}
\caption{Bulk flow magnitude and direction (Eqs.~(\ref{ucompress})
and (\ref{w_func1})) for the ENEAR, SN, A1SN and SFI++ catalogues:
(a) magnitude distribution; (b) 68\% contours for the bulk flow
direction, with $(\phi,\theta)=(l,\pi/2-b)$. Bulk flow directions
found by other probes and methods are also marked on the plot.}
\label{bffig1}
\end{figure*}

We now present our results on reconstructing bulk flow moments
using the minimum variance method on $50\hmpc$ scales
(Eq.~(\ref{ucompress})). In Fig.~\ref{bffig1}a, the magnitude of
the bulk flow is plotted for the four different catalogues.
Theoretically, the magnitude of bulk flow $v$ follows the
Maxwellian distribution, i.e.
\begin{equation}
p(V)dV=\sqrt{\frac{54}{\pi}} \left(\frac{V}{\sigma_{V}}
\right)^{2} \exp\left[ -\frac{3}{2} \left(\frac{V}{\sigma_{V}}
\right)^{2} \right] \frac{dV}{\sigma_{V}},
\end{equation}
where $\sigma_{V}$ is the velocity dispersion parameter
\citep{Coles02}. We calculate it as
$\sigma_{V}^{2}=\sigma^{2}_{x}+\sigma^{2}_{y}+\sigma^{2}_{z}=\sum_{i}C_{ii}$,
where $C$ is the covariance matrix (Eq.~(\ref{covar1})). One can
see that SFI++ provides the tightest constraint on the bulk flow
magnitude; this is because it is the largest, densest and closest
to full-sky survey available at the moment. In addition, ENEAR
(669 samples) and A1SN (153 samples) provide roughly similar
constraints on the bulk flow. This is due to the fact that
although the A1SN (First Amendment Supernovae catalogue) has less
data than ENEAR, its errors (calibrated by luminosity distance)
are much smaller than for the Fundamental Plane distance
estimates.

In Fig.~\ref{bffig1}a, one can also see that, although there are
offsets between the peaks of the likelihood for each individual
catalogue, they are all quite consistent with the theoretical
prediction, which is the mean-squared velocity of a $50\hmpc$
spherical region of $v\simeq 310 \kms$ (see
Section~\ref{mean_s_v}). Therefore by correcting the inhomogeneous
Malmquist bias and properly selecting the samples, the four
catalogues show a coherent flow on $50\hmpc$ scales of about
$310\kms$.

In addition, we plot the constraint on the direction of the bulk
flow in Fig.~\ref{bffig1}b, and compare these directions with
those found in other studies. From the figure, we can see that
SFI++ provides the tightest constraint on the bulk flow direction,
and the constraints on the direction of the bulk flows are
consistent with each other across all catalogues. We also mark the
preferred direction of the bulk flow from other published
estimates. We can see that our constraints are consistent with the
directions obtained from \citealt{Ma11}, \citealt{Nusser11a},
\citealt{Watkins09}, and \citealt{Macaulay12}, but
\cite{Kitaura12} prefer a slightly larger value for Galactic
latitude.

The quantitative results for the four catalogues are listed in
Table~\ref{tab2}.
\begin{table}
\begin{centering}
\begin{tabular}{@{}lccc}\hline
 & $v$ [$\times 100 \kms$]   & $l$ [degrees] & \ph{1}$b$ [degrees]\\
\noalign{\vspace{3pt}} \noalign{\hrule}
\ ENEAR & $2.2 \pm 0.6$ & $310 \pm 30$ & $\ph{1}-9.8 \pm 14$\\
\ A1SN &  $2.2 \pm 0.7$ & $290 \pm 60$ & $\ph{M\, }12.1 \pm 20$\\
\ SN &    $3.7 \pm 1.1$ & $290 \pm 30$ & $\ph{1}-0.7 \pm 15$\\
\ SFI++ & $3.4 \pm 0.4$ & $280 \pm \ph{0}8$ & $\ph{M1\, }5.1 \pm \ph{0}6$\\
\noalign{\vspace{-3pt}} \hline
\end{tabular}%
\caption{'Minimal variance' reconstructed magnitude and direction
(Eq.~(\ref{ucompress})) for the four catalogues. The quoted error
is the $\pm1\sigma$ measurement error.} \label{tab2}
\end{centering}
\end{table}

\subsection{Cosmological parameters}
\label{cosmologypara}

\begin{figure*}
\centerline{\includegraphics[bb=0 0 640
425,width=3.2in]{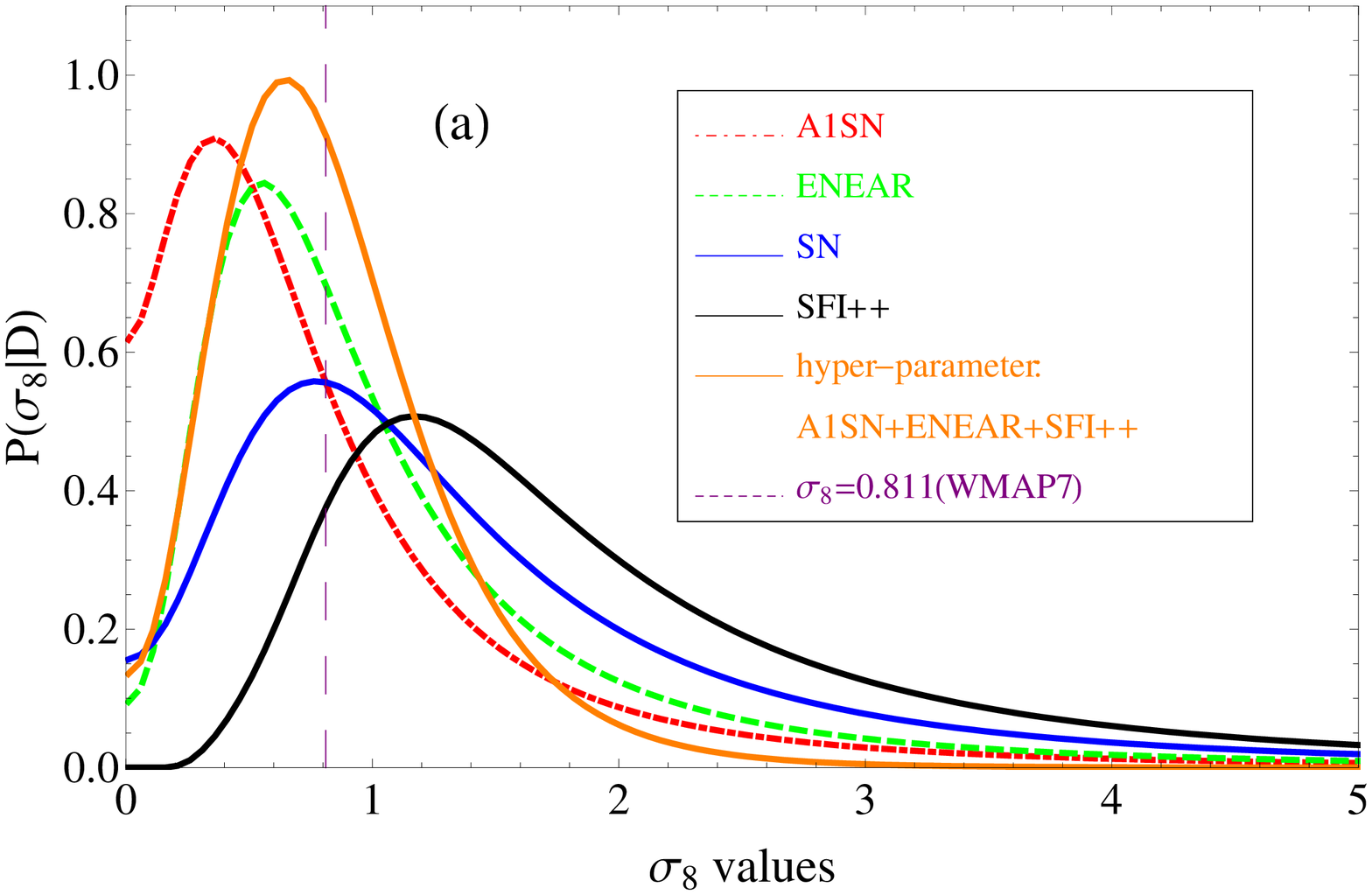}
\includegraphics[bb=0 0 480 480,width=2.8in]{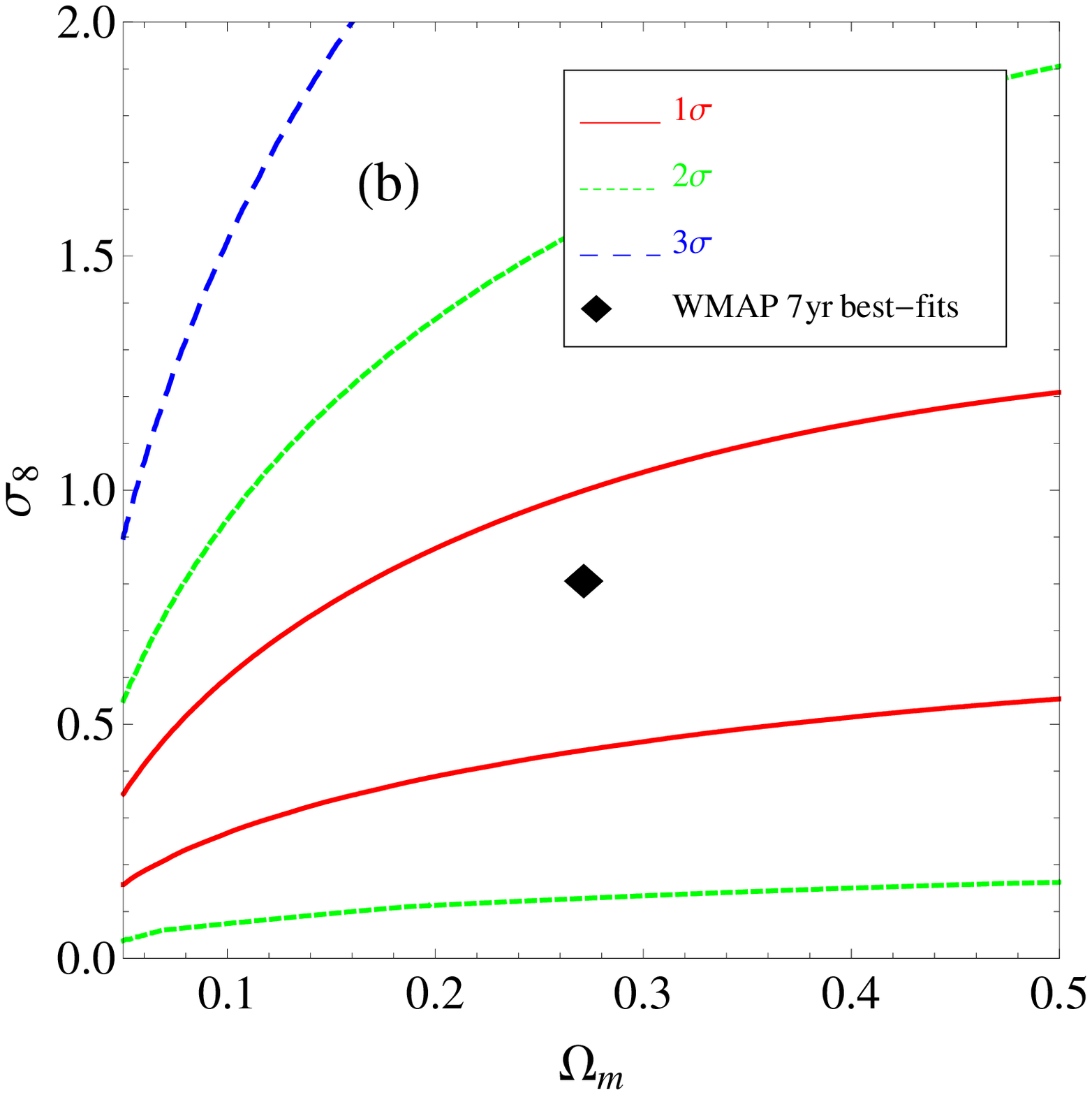}}
\caption{Cosmological parameter constraints: (a) the marginalised
distribution of $\sigma_{8}$, with the purple dashed vertical line
showing the {\it WMAP\/} 7-year best fit $\sigma_8$; (b) 2-D
contour plot in the $\sigma_{8}$--$\Omega_{\rm{m}}$ plane.}
\label{cpconstrain}
\end{figure*}

We now turn to cosmological parameter estimation. We first apply
the likelihood function (Eq.~(\ref{like1})) to to each individual
catalogue to calculate $P(\sigma_8|D)$, assuming a flat prior, and
then combine different catalogues by using the hyper-parameter
joint likelihood function of Eq.~(\ref{hyper_comb}). We show our
results in Fig.~\ref{cpconstrain}.

In Fig.~\ref{cpconstrain}a, one can see that the posterior
distribution for $\sigma_8$ is highly skewed and has a fairly long
tail out to large amplitudes, which suggests that the peculiar
velocity data available at the moment still cannot rule out flows
with large amplitude. In addition, we can see that the SN
catalogue peaks near the {\it WMAP\/} 7-year $\sigma_8$ value
($\sigma_8=0.811$), while the SFI++ catalogue prefers a slightly
higher value and A1SN and ENEAR prefer smaller ones. However,
within the errors they are all quite consistent with each other,
and none of them are inconsistent with the {\it WMAP\/} value of
$\sigma_8$.

In Fig.~\ref{cpconstrain}b, we plot the constraints on the
$\sigma_8$--$\Omega_{\rm{m}}$ plane by using the hyper-parameter
likelihood function of Eq.~(\ref{hyper_comb}). One can see that
the {\it WMAP\/} best-fit value is located close to the 68\%
contour in the $\sigma_8$--$\Omega_{\rm{m}}$ plane, and therefore
the hyper-parameter results are consistent with the expectation
from $\Lambda$CDM. Comparing Fig.~\ref{cpconstrain}b with figure 6
in \cite{Watkins09}, one can see that our contour prefers a much
lower value of $\sigma_{8}$, and it is also closer to the {\it
WMAP} value of $\Omega_{\rm{m}}$.

\subsection{Multi-shells likelihood method}
\label{correlation-depth}

\begin{figure*}
\centerline{\includegraphics[bb=0 0 477
323,width=3.0in]{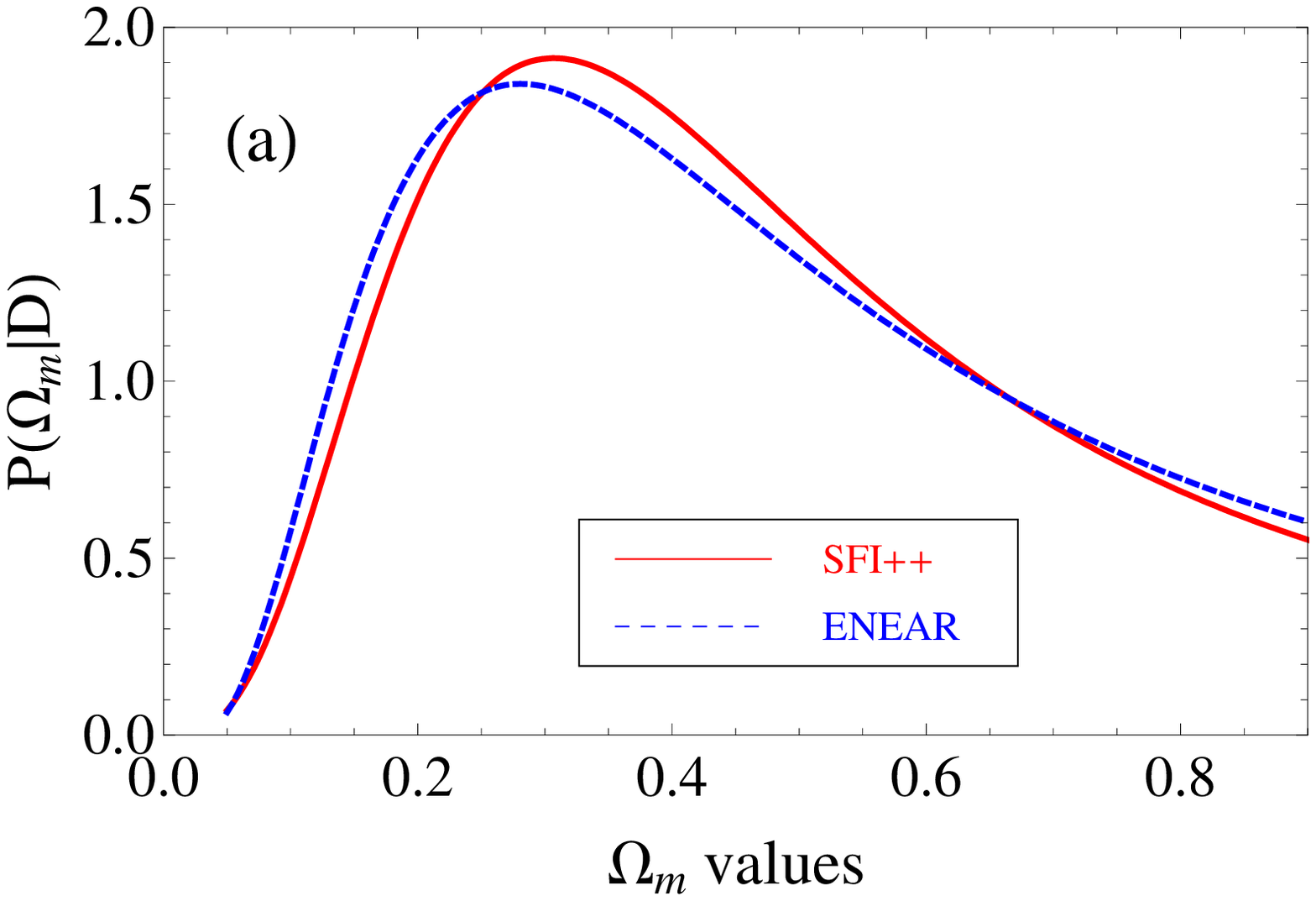}
\includegraphics[bb=0 0 494 334,
width=3.0in]{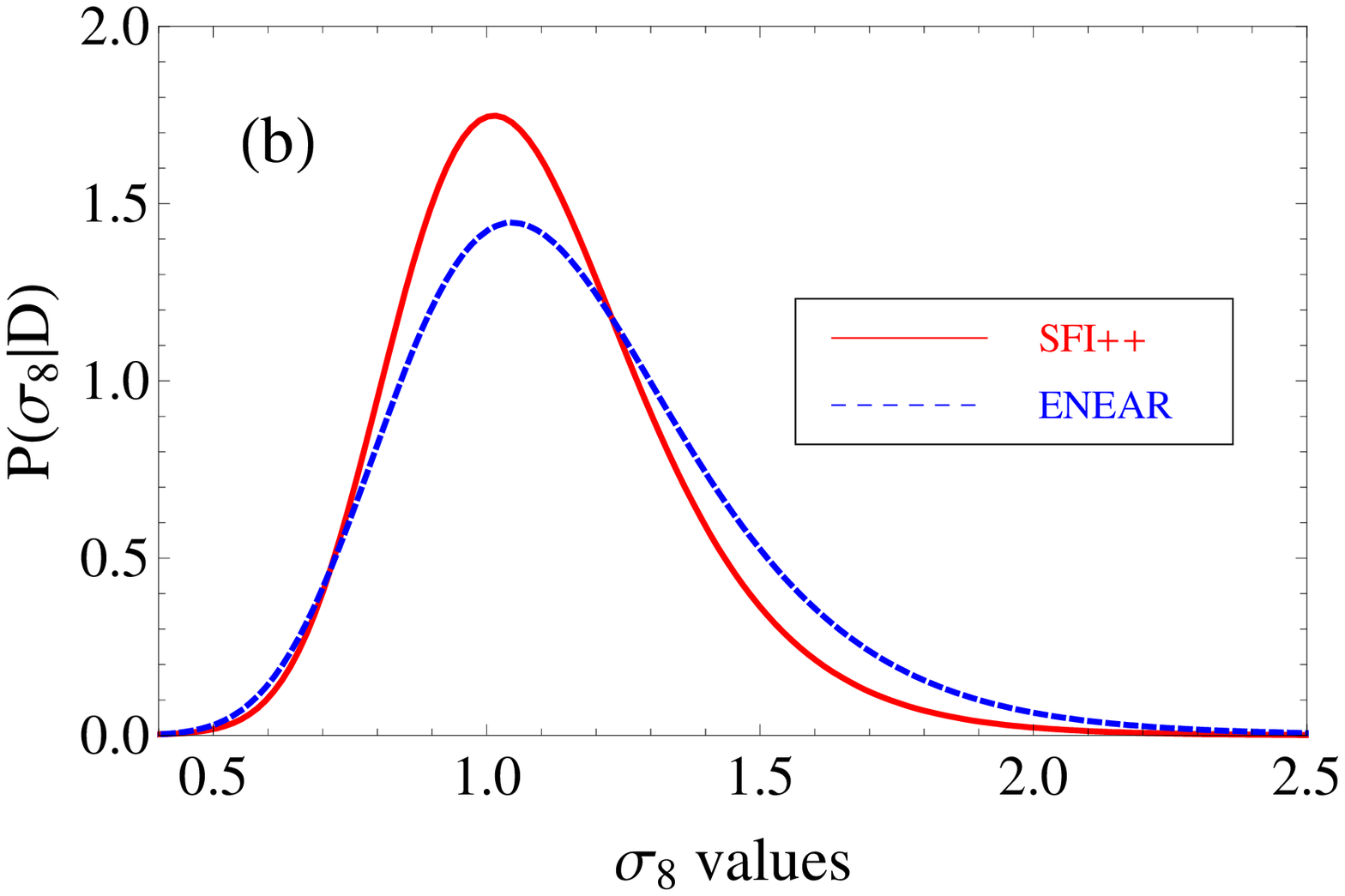}} \centerline{\includegraphics[bb=0 0
485 485,width=3.0in]{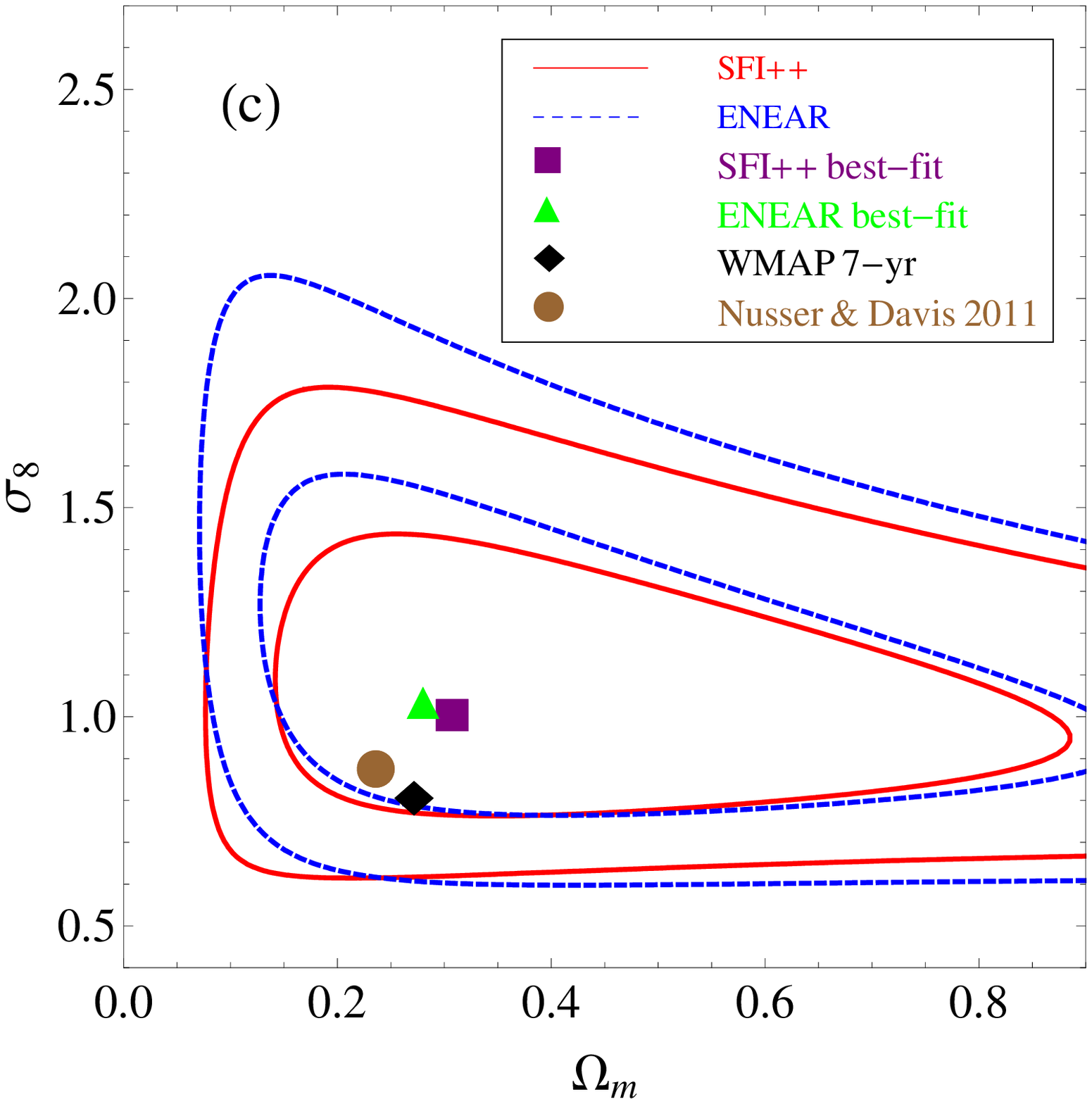}} \caption{Marginalised
distributions of $\sigma_{8}$ (panel (a)) and $\Omega_{\rm{m}}$
(panel (b)) parameters from the SFI++ and ENEAR catalogues, by
using the likelihood function (Eq.~(\ref{like3})) for $8$
correlated shells with distances $20$--$90 \hmpc$. Panel (c) shows
the 2-D contours of the joint distribution
$\sigma_{8}$--$\Omega_{\rm{m}}$. The {\it WMAP} 7-yr best-fit
values and results from Nusser et al. (2011) are also plotted.}
\label{fig:sig8omg:shell}
\end{figure*}

\begin{table*}
\begin{centering}
\begin{tabular}{@{}lccc}\hline
 & $\Omega_{\rm{m}}$  &  $\sigma_{8}$ & references \\
\noalign{\vspace{2pt}} \noalign{\hrule} \noalign{\vspace{3pt}}
\ SFI++ multishells & $\phantom{0}0.31^{+0.28}_{-0.14}$ & $1.01^{+0.26}_{-0.20}$ & This study \\
\ ENEAR multishells & $\phantom{0}0.28^{+0.30}_{-0.14}$ & $1.04^{+0.32}_{-0.24}$ & This study \\
\ {\it WMAP} 7-year & $\phantom{0}0.271 \pm 0.016$ & $0.811^{+0.030}_{-0.031}$ & \cite{Komatsu11} \\
\ SFI++ ASCE method & $0.235^{+0.16}_{-0.09}$ & $0.86 \pm 0.11$& \cite{Nusser11a} \\
\noalign{\vspace{-3pt}} \hline
\end{tabular}%
\caption{Comparison on the constraints on cosmological parameters
$\Omega_{\rm{m}}$ and $\sigma_{8}$ from the multishell likelihood
(Eq.~(\ref{like3})) and two other probes. Since Nusser et al.\
(2011a) do not explicitly quote the errors of the parameters, we
roughly estimate the constraints from their figures 9 and 10.}
\label{tab:comp1}
\end{centering}
\end{table*}

In the above approach, we use the reconstructed 3-D bulk flow
velocity as the `observational data' to constrain cosmology. We
have shown that this likelihood (Eq.~(\ref{like1})) can provide a
fairly strong constraint on $\sigma_{8}$, but the constraint on
$\Omega_{\rm{m}}$ is rather weak and thus the 2-D contours of
$\sigma_{8}$--$\Omega_{\rm{m}}$ do not close. This is because we
use only one velocity vector (at $50 \hmpc$) as a constraint and
this information is not enough to provide a tight limit (see also
figure $6$ in \citealt{Watkins09}).

Based on the `minimal variance' scheme, here we propose another
method to consider the bulk flow velocities for {\it all\/} of the
shells within a certain radius. Since bulk flow velocities on
different shells are highly correlated, one needs to calculate the
full-covariance matrix of those bulk flow velocities. Note that we
will apply this method to each individual data set to assess its
validity for constraining cosmological parameters.

In the step of calculating $\langle S_{j}U_{p} \rangle$
(Eq.~(\ref{vel_smup1})), we need to simulate $N'=10^{4}$ random
velocities in the top-hat region $R$ and therefore obtain the
weighting function $w_{i,n}$ for a certain shell $R$. Let us
assume we can sample multiple shells with this method, and
therefore obtain the weighting function $w^{R}_{i,n}$ and bulk
flow velocity $u^{R}_{i}$ for each shell $R$. Now we can calculate
the covariance matrix of $u^{R}_{i}$s as ($R,R'$ are two shells)
\begin{eqnarray}
C^{RR'}_{pq} &=& \langle u^{R}_{p}u^{R'}_{q} \rangle  \notag \\
&=& \sum_{i,j}w^{R}_{p,i}w^{R'}_{q,j}\langle S_{i} S_{j} \rangle  \notag \\
&=&\sum_{i,j}w^{R}_{p,i}w^{R'}_{q,j} G_{ij}, \label{covar2}
\end{eqnarray}
which can be broken down into an instrumental noise term
\begin{equation}
C^{{\rm n},RR'}_{pq}=
 \sum_{i}w^{R}_{p,i}w^{R'}_{q,i}\left(\sigma^{2}_{i}+\sigma^{2}_{\ast}
\right),  \label{noise_mat2}
\end{equation}
and a cosmic variance term
\begin{equation}
C^{{\rm v},RR'}_{pq}=\frac{\Omega^{1.1}_{\rm m} H^{2}_{0}}{2
\pi^{2}} \int dk \text{ }P(k) \text{
}\left(W^{RR'}_{pq}\right)^{2}(k),
\end{equation}
where the angle-averaged window function is
\begin{equation}
\left(W^{RR'}_{pq}\right)^{2}(k)=\sum_{i,j}w^{R}_{p,i}w^{R'}_{q,j}F_{ij}(k).
\label{vel_w2pq2}
\end{equation}
Note that all of the shells are correlated. Suppose we have $M$
shells, and now we arrange the bulk flow velocities of each shell
into a $3\times M$ velocity vector $\tilde{u}_{i}$, where $i$ runs
from $1$ to $3M$. For instance, the $y$-direction of bulk flow in
shell $3$ is now at the position $3\times(3-1)+2=8$ in the
$\tilde{u}_{i}$ vector. We do the same thing for the covariance
matrix, and therefore we turn the covariance matrix $C^{RR'}_{pq}$
into a $3M \times 3M$ covariance matrix $\Sigma=\Sigma^{\rm
v}+\Sigma^{\rm n}$, where the $\Sigma^{\rm v}$ part contains the
cosmological parameters and $\Sigma^{\rm n}$ includes measurement
errors. Now the likelihood function for multiple shells becomes
\begin{equation}
\mathcal{L}(\mathbf{\theta})= \frac{1}{(2 \pi)^{\frac{3}{2}}
 \det(\Sigma(\mathbf{\theta}))^{\frac{1}{2}}}\exp
   \left(-\frac{1}{2} \sum^{3M}_{i,j=1}u_{i}\Sigma(\mathbf{\theta})^{-1}_{ij}u_{j}
   \right). \label{like3}
\end{equation}

We apply this likelihood function to the SFI++ and ENEAR
catalogues, since they are the deeper catalogues with the broader
sky-coverage. The SFI++ catalogue has a mean distance of
${\sim}\,40 \hmpc$ and extends out to $180 \hmpc$, whereas the
ENEAR catalogue has a mean distance ${\sim}\,30 \hmpc$ and goes
out to $150 \hmpc$. We first trim both data sets out to $100
\hmpc$, which leaves $2830$ (SFI++) and $690$ (ENEAR) samples.
Then we calculate the weighting functions $w_{i,n}$ and bulk flow
velocities $u_{i}$ for $8$ different shells of distances $20$--$90
\hmpc$, each with $10 \hmpc$ separation. The reason we use the
bulk flows only on shells with distances greater than $20 \hmpc$
is that we would like to avoid non-linear structures on small
scales. In addition, more distant objects are not very well
sampled and therefore are very sparse, so we restrict our bulk
flows to within the shell of $90\hmpc$. During the process of
computation, we stick to the same conventions as listed in
Section~\ref{individual_like}. Then we calculate the covariance
matrix (Eq.~(\ref{covar2})) and the likelihood function
(Eq.~(\ref{like3})) for the $8$ shells, and we obtain the
marginalised distribution of $\Omega_{\rm{m}}$ and $\sigma_{8}$,
as shown in Figs.~\ref{fig:sig8omg:shell}a and
\ref{fig:sig8omg:shell}b. The joint distribution of
$\sigma_{8}$--$\Omega_{\rm{m}}$ is shown in
Fig.~\ref{fig:sig8omg:shell}c.

From Fig.~\ref{fig:sig8omg:shell}a and
Fig.~\ref{fig:sig8omg:shell}c, one can see that the constraint on
$\Omega_{\rm{m}}$ becomes tighter than the previous single bulk
flow constraint and its $1\sigma$ contour is now closed.
Therefore, by just using bulk flow data, one can obtain an
independent constraint on the cosmological parameters. The
best-fit value of {\it WMAP} 7-year results, as well as the
constraints obtained from \cite{Nusser11a} are all well within the
$\pm 1\sigma$ confidence region of the parameter space. The reason
that the likelihood function for multiple shells can give a
reasonably good constraint on $\Omega_{\rm{m}}$, while the bulk
flow at $50 \hmpc$ does not, is that the the dependence of $P(k)$
on $\Omega_{\rm{m}}$ is a function of scale, and therefore by
incorporating multiple shells, one can gain more information on
perturbations at different depths.

We would also like to point out that since the current peculiar
velocity data are no deeper than $150\hmpc$, and the data beyond
$100 \hmpc$ are very noisy and sparse, the $8$ shell bulk flows at
distances of $20$ to $90 \hmpc$ are really the maximal information
we can obtain with these catalogues. We have carefully checked
that, for the data within $100 \hmpc$, splitting into more shells
of bulk flows does not improve the constraints, since these shells
are highly correlated and we already have enough shells to
effectively capture the scale dependence.

In addition, we should note that there is another statistical
approach, the multiple moment method
\citep{Jaffe95,Feldman10,Macaulay11,Macaulay12}, which has been
proposed to reconstruct the bulk flow, shear, and octupole moments
of perturbations. This can be considered as an alternative method
to our multishell likelihood approach.  The multiple moment method
used in \citet{Feldman10} and \citet{Macaulay11} considers
perturbations only on $50 \hmpc$ scales, but includes all moments,
and they find that there is excessive power for the bulk flow, but
not for the other moments. In contrast, our multishell likelihood
function focuses just on the bulk flow, and it reconstructs this
for shells of different distance, quantifying the full covariance
matrix by calculating the correlations between shells. Our
multishell likelihood shows that the bulk flow is not excessive
compared with $\Lambda$CDM predictions, and that one can obtain
reliable constraints on cosmological parameters by applying the
method to various peculiar velocity catalogues.

We list the numerical results of our cosmological parameter
constraints in Table~\ref{tab:comp1}.

\section{Discussion and Conclusions}
\label{vel_conclude}

\begin{table*}
\begin{centering}
\begin{tabular}{@{}lll}\hline
 & \cite{Watkins09} & This study \\
\noalign{\vspace{3pt}} \noalign{\hrule} \noalign{\vspace{3pt}}
Cosmological parameters & {\it WMAP\/} 5-year \citep{Komatsu09} &
 {\it WMAP\/} 7-year \citep{Komatsu11} \\
Small scale velocity dispersion $\sigma_{\ast}$ & $150 \kms$ &
$400 \kms$
\\
$P(k)$ calculation & Parameterisation with
$\Gamma=\Omega_{\rm{m}}h$ & Numerical result from CAMB
\\
Distance indicator & Malmquist bias uncorrected & Malmquist bias corrected \\
Catalogues & COMPOSITE: combination of
& SN, SFI++ and ENEAR catalogues \\
& SBF, SN, SFI++, ENEAR& \\
& SC, SMAC, EFAR and Willick  & \\
Data selection & None & Trim to $d \leq 80\hmpc$, $|v| \leq 3000 \kms$ \\
Number of samples & COMPOSITE (4536) & SN (78), ENEAR (669), SFI++ (2404) \\
Combination method & Direct combination & Hyper-parameter
 likelihood $\&$ Multi-shell likelihood \\
\hline \textbf{Result for normalisation} & $\sigma_{8}=1.7 \pm
0.28$  & $\sigma_{8}=0.65^{+0.47}_{-0.35}$ (hyper-parameter) \\
& & $\sigma_{8}=1.01^{+0.26}_{-0.20}$ (SFI++) \text{ }
$\sigma_{8}=1.04^{+0.32}_{-0.24}$ (ENEAR) \\
& (excluded by {\it WMAP\/} at $99\%$) & (consistent with
{\it WMAP\/}) \\
\hline
\end{tabular}%
\caption{Comparison of the methodology, data selection, and
results of our constraints with those in \citet{Watkins09}.}
\label{tab3}
\end{centering}
\end{table*}

In this paper, we have been investigating bulk flow measurements
using various catalogues. We find results which are different to
those given by \cite{Watkins09}, who claimed evidence for a
surprisingly large bulk flow on $50\hmpc$ scales, apparently
discrepant with the $\Lambda$CDM prediction. In contrast, by
carefully considering four selected catalogues, we find a coherent
flow of about $300 \kms$ on a scale of $50\hmpc$, entirely
consistent with the value expected given the {\it WMAP\/} 7-year
cosmological parameters.

By employing the same weighting scheme and the same conventions,
we are able to accurately reproduce the results in
\cite{Watkins09}, as shown in Fig.~\ref{reproduce1}.  Since we
focus on the SN, SFI++ and ENEAR catalogues, we removed the other
sub-catalogues from the COMPOSITE catalogue, and found a slightly
lower value of $\sigma_8$ (red line in Fig.~\ref{reproduce1}b),
but still higher than the {\it WMAP\/} constraint. This indicates
that the high value of $\sigma_8$ inferred from the COMPOSITE
catalogue is not completely driven by the five deep and sparse
catalogues included (SMAC, SBF, SC, EFAR and Willick).

To summarise the various other issues which could be responsible
for the discrepancy, in Table~\ref{tab3} we list several technical
points which lead to quantitatively different results.

The first issue is the assumption of small-scale velocity
dispersion, which goes into the calculation of the covariance
matrix (Eq.~(\ref{noise_mat1})). \cite{Watkins09} assumed a value
of $150 \kms$, which is too small compared to the constraint
obtained by \cite{Ma11,Ma11b}, which was closer to the $400\kms$
we chose here.  Besides this, \cite{Watkins09} used an inaccurate
approximation for the matter power spectrum. From
Fig.~\ref{pkcompare1}, one can see that although this is a small
effect, it has the same sign, yielding smaller flows. Thus, by
fitting to the observed flows, this tends to further increase the
normalisation parameter $\sigma_{8}$.

The second major difference lies in the inhomogeneous Malmquist
bias correction. In \cite{Watkins09}, only the SFI++ and SMAC
catalogues were corrected for this effect.  In our approach, we
used the full-sky density field from the PSCz catalogue to
extrapolate the density $n(r)$ at any spatial position, and
calculate the probability of the true distance $r$ given the
measured distance $d$ (Eq.~(\ref{MBP1})).  The comparison between
the measured distance/velocity and true distance/velocity in
Fig.~\ref{MBcorrect}, shows that the bias tends to move galaxies
to smaller distances.

Another difference is that we only keep the high quality samples
SN, SFI++ and ENEAR from the \cite{Watkins09} compilation, and we
further include the recent compilation of supernovae data, i.e.\
the A1SN catalogue. To remove any possible bias from the distant
and sparsely sampled region, we restricted our attention to $d
\leq 80\hmpc$, and to avoid the results being driven by outliers,
we also limited our samples to $|v|\leq 3000 \kms$.

Furthermore, rather than using the COMPOSITE catalogue, we
combined individual sample likelihoods using the Bayesian
hyper-parameter technique. This should avoid the possibility that
inconsistent data sets may bias the result if they are assigned
equal weight. From the hyper-parameter likelihood, we find the
best-fit value $\sigma_8=0.65^{+0.47}_{-0.35}$. This is somewhat
low and hence inconsistent with a large bulk flow. However, the
uncertainty is so large that this result is still consistent with
standard $\Lambda$CDM expectations.

Finally, we proposed a multishell likelihood method, which
calculates the bulk flows in all shells within a certain radius
together with their covariance matrix. This multishell likelihood
takes into account the scale-dependence of the matter power
spectrum $P(k)$ on the $\Omega_{\rm{m}}$ parameter, and therefore
maximises the constraining power one can obtain from a data set.
By applying this likelihood to the SFI++ and ENEAR catalogues, we
showed that they can provide much stronger constraints on
$\Omega_{\rm{m}}$ and $\sigma_{8}$ than the single shell ($50
\hmpc$) constraint. Our result also shows consistency with {\it
WMAP} 7-year best-fits and results from \cite{Nusser11a}.

We conclude that the apparently large bulk flow on $50\hmpc$
scales found by \cite{Watkins09} may not be a genuine flow. By
correcting for Malmquist bias, carefully selecting samples and
examining assumptions, one finds that the current peculiar
velocity field catalogues are consistent with the $\Lambda$CDM
model. On the other hand, any claimed discrepancy is not due to
the `minimal variance' scheme proposed by \cite{Watkins09} and
\cite{Feldman10}, since in our tests, we have shown that this
scheme gives consistent results. In addition, our conclusions also
agree with several other independent searches for bulk flows, such
as the ASCE method with the SFI++ catalogue \citep{Nusser11a}, the
minimal variance method with the Type-Ia SN data
\citep{Turnbull12}, and the luminosity function method with the
2MRS samples \citep{Branchini12}. It should also be pointed out
that the lack of evidence for a bulk flow on $50 \hmpc$ removes
some of the support for an excessive flow $\sim1000 \kms$ on even
deeper scales $\sim300 \hmpc$ \citep{Kashlinsky08}.

It seems clear that, despite extensive effort for decades,
peculiar velocity catalogues remain systematics dominated. By
applying different, but apparently reasonable, assumptions and
statistical approaches, it is possible to find quite discrepant
results using essentially the same data sets.  This means that the
realistic error bars are probably larger than given in many of the
published studies. In addition, one should notice that there are
many other methods developed to compute bulk flows that do not
rely on distance indicators, such as luminosity fluctuations and
fluctuations in the galaxy number density \citep{Branchini12}, as
well as the use of the kinetic Sunyaev-Zeldovich effect (e.g.
\citealt{Osborne11}). Although they also suffer from systematic
effects, these will be of a different nature and therefore such
approaches can be regarded as complementary to the method
discussed here. Large-scale bulk flows still offer promise for
constraining cosmological models, but fully realising that promise
will require further improvements in the construction of
catalogues, and in the control of the systematic effects which
continue to plague this field.

\vskip 0.1 truein

\noindent \textbf{Acknowledgments:} We would like to thank Michael
Hudson and Stephen Turnbull for sharing with us the First
Amendment Supernovae compilation, and Enzo Branchini and George
Efstathiou for helpful discussions. This research was supported by
the Natural Sciences and Engineering Research Council of Canada.
YZM is supported by a CITA National Fellowship.

\appendix

\section{Power spectrum Semi-analytic formula analysis}
\label{powerspectrum_form}
\begin{figure*}
\centerline{\includegraphics[bb=0 0 536
336,width=3.4in]{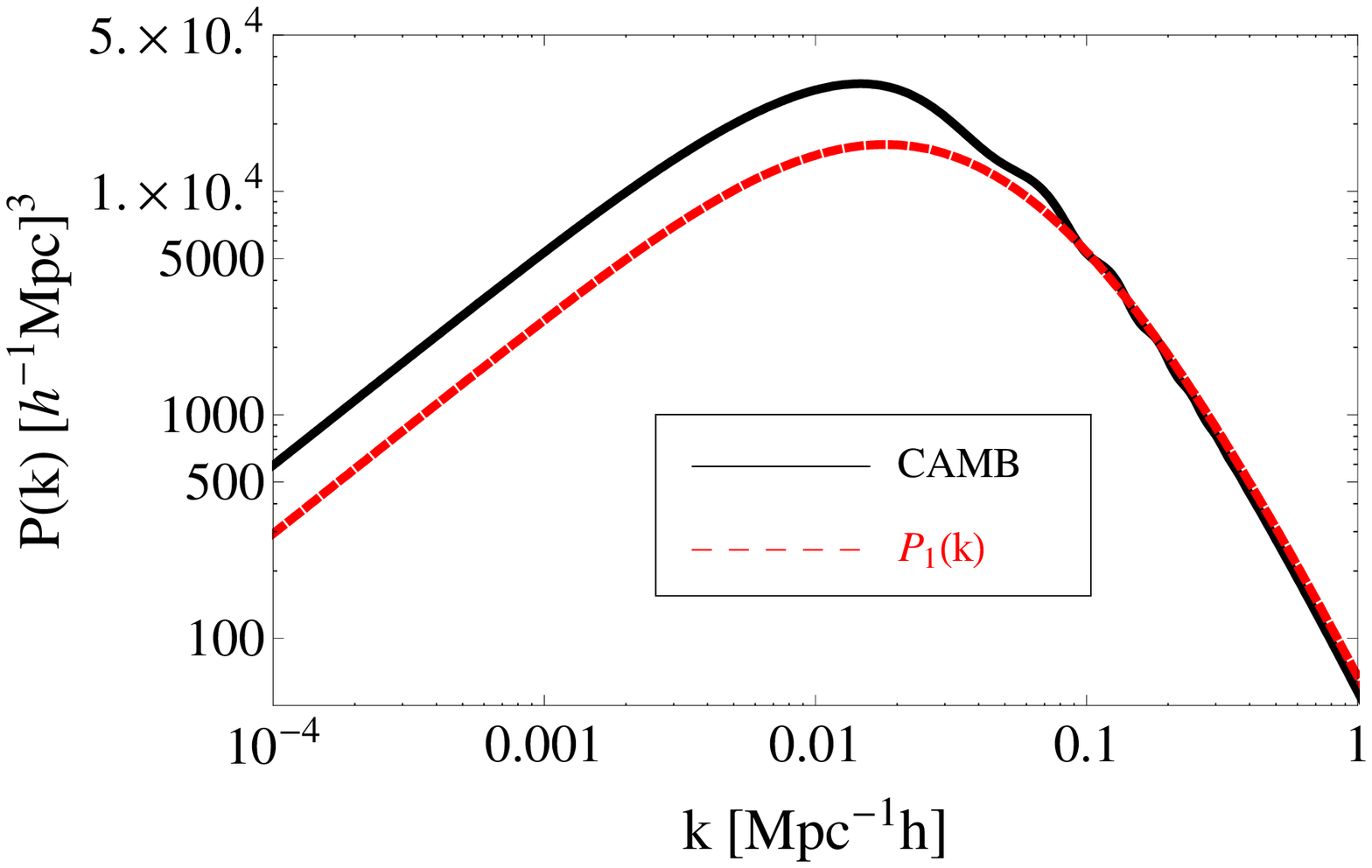}
\includegraphics[bb=0 0 593 391,width=3.2in]{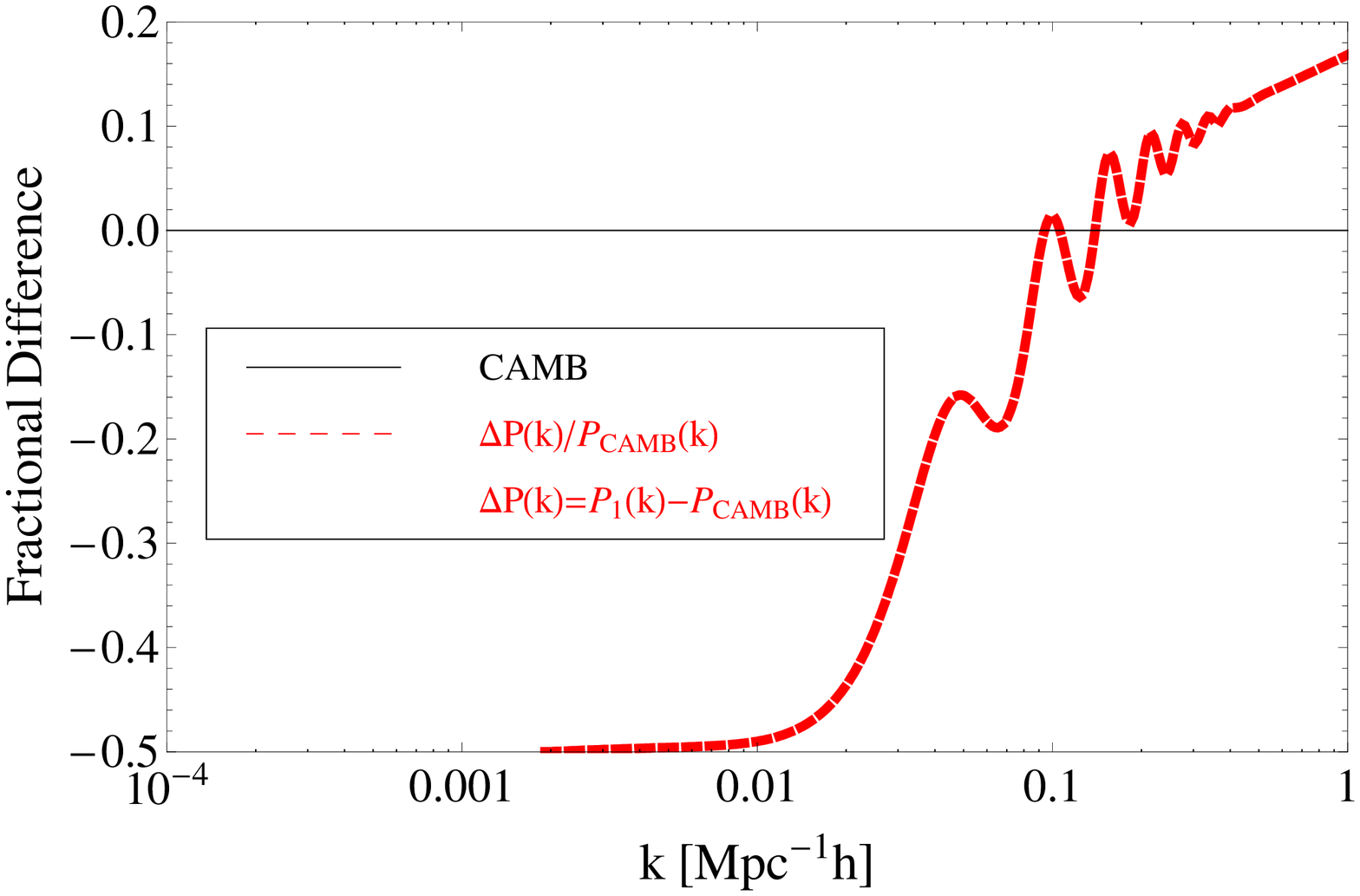}}
\caption{Comparison of the approximation $P_{1}(k)$ using shape
parameter $\Gamma=\Omega_{\rm{m}}h$ with the numerical result from
CAMB \citep{camb}. Here we adopt {\it WMAP\/} 5-year best-fit
parameters \citep{Komatsu09}. One can see that there are large
deviations on scales of $10^{-4}$ to $0.1 \mpch$.}
\label{pkcompare1}
\end{figure*}

To sample the $\sigma_{8}$--$\Omega_{\rm{m}}$ parameter space, we
can apply a formula to generate the matter power spectrum $P(k)$.
We use the following semi-analytic equation as presented in
\cite{Eisenstein98}:

\begin{equation}
P(k)=\sigma_{8}^{2}k^{n_{\rm s}}T(k)^{2},
\end{equation}%
where
\begin{eqnarray}
T(k) =\frac{L_{0}}{L_{0}+C_{0}(k,\Gamma )\left(k/\Gamma
\right)^{2}},\notag \\
L_{0} =\ln (2e+1.8\left(k/\Gamma \right) ),  \notag \\
C_{0}(k,\Gamma ) =14.2+\frac{731}{1+62.5\left(k/\Gamma \right) }.
\end{eqnarray}
The quantity $\Gamma$ here is called the power spectrum `shape
parameter'. \cite{Watkins09} and \cite{Feldman10} used
$\Gamma=\Omega_{\rm{m}}h$ as an approximation on large scales.
However, in Fig.~\ref{pkcompare1}, we can see that this
approximation ($P_1(k)$) still has relatively large deviations
from the numerical result from {\sc camb}.

A more accurate parameterisation of the shape parameter is
$\Gamma=\Omega_{\rm{m}}h \exp(-\Omega_{\rm
b}(1+\sqrt{2h}/\Omega_{\rm{m}}))$, as advocated in
\cite{Eisenstein98}. This is much closer to the numerical $P(k)$,
due to the additional exponential factor.

In order to demonstrate the successful reproduction of results in
\cite{Watkins09}, we use the approximation
$\Gamma=\Omega_{\rm{m}}h$ in Section~\ref{mv_scheme}. However,
since in our subsequent analysis, we need to carefully compare the
numerical value of the reconstructed bulk flow moment with the
expectation based on cosmological parameters, we switch to the
numerical result of the $P(k)$ from CAMB \citep{camb} in our
determination of the bulk flow moments in each individual
catalogue and subsequent hyper-parameter analysis.

\end{document}